\documentclass[twocolumn,amsmath,aps,prd,preprintnumbers,amssymb,nofootinbib,superscriptaddress,showpacs,floatfix]{revtex4-1}
\usepackage{amssymb}
\usepackage{amsmath}
\usepackage{epsfig}
\usepackage{subfigure}
\usepackage{mathrsfs}
\usepackage{hyperref}
\usepackage{longtable}
\usepackage[usenames,dvipsnames]{xcolor}
\usepackage{mathtools}
\usepackage{relsize}

\begin{document}
 \renewcommand{\thefigure}{\arabic{figure}}
\newcommand{\noj}{}

\newcommand{\apjl}{Astrophys. J. Lett.}
\newcommand{\aap}{Astron. Astrophys.}
\newcommand{\apjs}{Astrophys. J. Suppl. Ser.}
\newcommand{\sa}{Sov. Astron. Lett.}.
\newcommand{\jpb}{J. Phys. B.}
\newcommand{\natu}{Nature (London)}
\newcommand{\aaps}{Astron. Astrophys. Supp. Ser.}
\newcommand{\aj}{Astron. J.}
\newcommand{\aas}{Bull. Am. Astron. Soc.}
\newcommand{\mnras}{Mon. Not. R. Astron. Soc.}
\newcommand{\pasp}{Publ. Astron. Soc. Pac.}
\newcommand{\jcap}{JCAP.}
\newcommand{\jmat}{J. Math. Phys.}
\newcommand{\prep}{Phys. Rep.}
\newcommand{\jtep}{Sov. Phys. JETP.}
\newcommand{\plb}{Phys. Lett. B.}
\newcommand{\pla}{Phys. Lett. A.}
\newcommand{\jhep}{Journal of High Energy Physics}

\newcommand{\be}{\begin{equation}}
\newcommand{\ee}{\end{equation}}
\newcommand{\bea}{\begin{align}}
\newcommand{\eea}{\end{align}}
\newcommand{\lsim}{\mathrel{\hbox{\rlap{\lower.55ex\hbox{$\sim$}} \kern-.3em \raise.4ex \hbox{$<$}}}}
\newcommand{\gsim}{\mathrel{\hbox{\rlap{\lower.55ex\hbox{$\sim$}} \kern-.3em \raise.4ex \hbox{$>$}}}}
\newcommand{\grad}{\ensuremath{\vec{\nabla}}}
\newcommand{\adotoa}{\ensuremath{{\cal H}}} 
\newcommand{\Uc}{\ensuremath{{\cal U}}}
\newcommand{\Vc}{\ensuremath{{\cal V}}}
\newcommand{\Jc}{\ensuremath{{\cal J}}}
\newcommand{\Mc}{\ensuremath{{\cal M}}}

\newcommand{\unit}[1]{\ensuremath{\, \mathrm{#1}}}

\newcommand{\tkDM}[1]{\textcolor{red}{#1}}              \newcommand{\tkRH}[1]{\textcolor{blue}{#1}}

\title{Future CMB tests of dark matter: ultra-light axions and massive neutrinos}

\author{Ren\'{e}e Hlo\v{z}ek}
\email{hlozek@dunlap.utoronto.ca}
\affiliation{Dunlap Institute for Astronomy and Astrophysics, University of Toronto, 50 St. George Street, Toronto, Ontario, Canada M5S 3H4}
\affiliation{Department of Astronomy and Astrophysics, University of Toronto, 50 St. George Street, Toronto, Ontario, Canada M5S 3H4}
\author{David J.~E.~Marsh}
\email{david.marsh@kcl.ac.uk}
\affiliation{King's College London, Strand, London, WC2R 2LS, United Kingdom}
\author{Daniel Grin}
\email{dgrin@kicp.uchicago.edu}
\affiliation{Kavli Institute for Cosmological Physics, Department of Astronomy and Astrophysics, University of Chicago, Chicago, Illinois, 60637, U.S.A.}
\author{Rupert Allison}
\affiliation{University of Oxford, Denys Wilkinson Building, Keble Road, OX1 3RH}
\author{Jo Dunkley}
\affiliation{University of Oxford, Denys Wilkinson Building, Keble Road, OX1 3RH}
\author{Erminia Calabrese}
\affiliation{University of Oxford, Denys Wilkinson Building, Keble Road, OX1 3RH}
\email{}\date{\today}
\begin{abstract}
Measurements of cosmic microwave background (CMB) anisotropies provide strong evidence for the existence of dark matter and dark energy. They can also test its composition, probing the energy density and particle mass of different dark-matter and dark-energy components. CMB data have already shown that ultra-light axions (ULAs) with mass in the range $10^{-32}~{\rm eV} \to 10^{-26}~{\rm eV}$ compose a fraction $\lsim 0.01$ of the cosmological critical density. Here, the sensitivity of a
proposed CMB-Stage IV (CMB-S4) experiment (assuming a 1 arcmin beam and $\sim 1~\mu K{\rm-arcmin}$ noise levels over a sky fraction of 0.4) to the density of ULAs and other dark-sector components is assessed. CMB-S4 data should be $\sim 10$ times more sensitive to the ULA energy-density than {\sl Planck} data alone, across a wide range of ULA masses $10^{-32}\lsim m_{a}\lsim 10^{-23}~{\rm eV}$, and will probe axion decay constants of $f_{a}\approx 10^{16}~{\rm GeV}$, at the grand unified scale. CMB-S4 could improve the CMB lower bound on the ULA mass from $\sim 10^{-25}~{\rm eV}$ to $10^{-23}~{\rm eV}$, nearing the mass range probed by dwarf galaxy abundances and dark-matter halo density profiles. These improvements will allow for a multi-$\sigma$ detection of percent-level departures from CDM over a wide range of masses. Much of this improvement is driven by the effects of weak gravitational lensing on the CMB, which breaks degeneracies between ULAs and neutrinos. We also find that the addition of ULA parameters does not significantly degrade the sensitivity of the CMB to neutrino masses. These results were obtained using the \textsc{axionCAMB} code (a modification to the \textsc{CAMB} Boltzmann code), presented here for public use.

\end{abstract}
\maketitle

\section{Introduction}
\label{intro}

\begin{figure}[htbp!] 
\includegraphics[width=0.5\textwidth, trim = 5mm 0mm 10mm 10mm, clip]{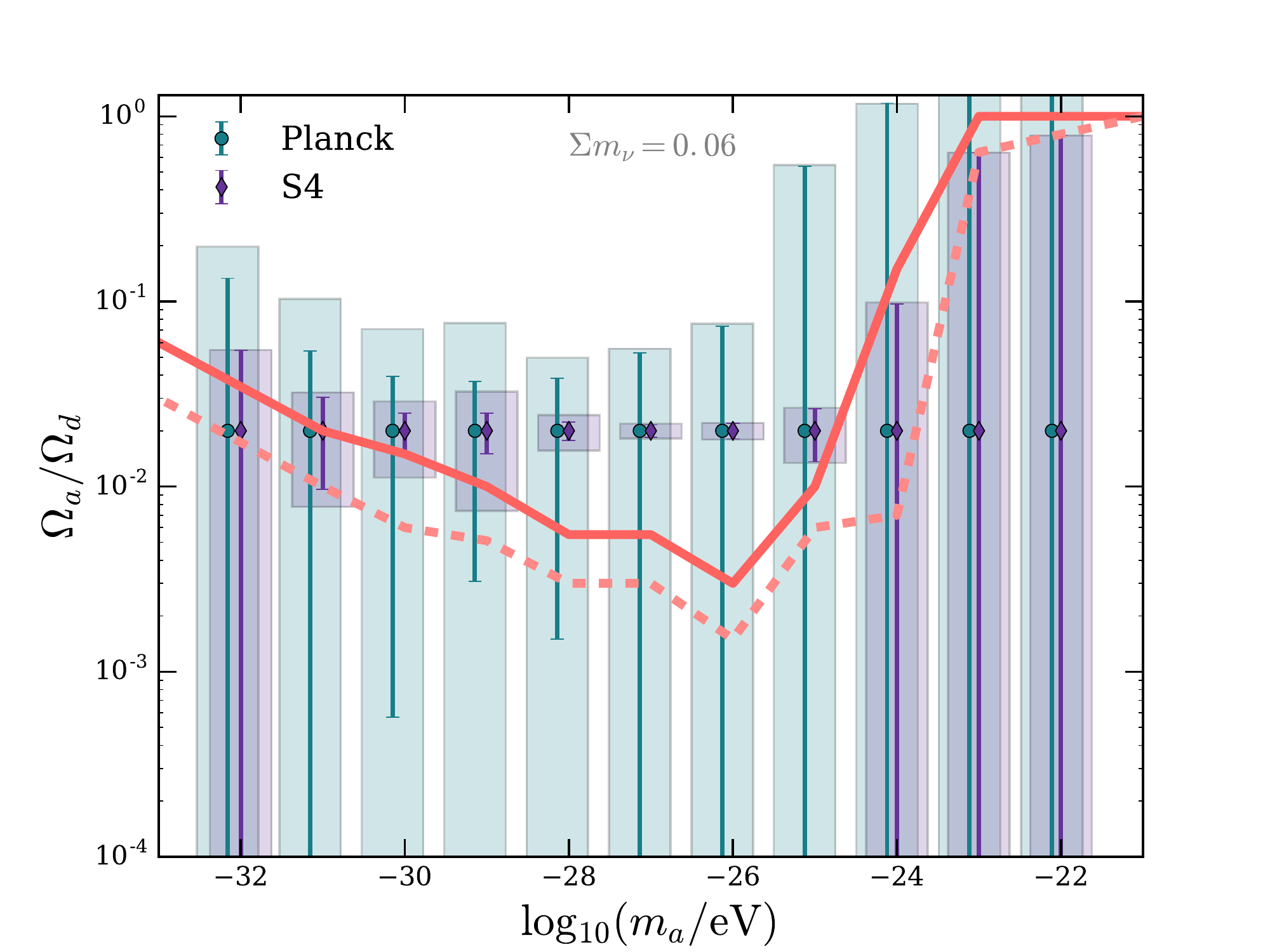}
\caption{\textbf{Projected CMB-S4 sensitivity to the axion energy density as a function of axion mass, compared with Fisher-matrix {\sl Planck} sensitivity}: Vertical bars show $1\sigma$ errors at fixed neutrino mass $\Sigma m_\nu = 0.06~\mathrm{eV}$ while the shaded bars show the errors marginalizing over $\Sigma m_{\nu}$. We classify axions as DE-like if $m_a < 10^{-29}\,\mathrm{eV}$, `DM-like' if $m_a > 10^{-25} \mathrm{eV}$ and `fuzzy' DM for masses in between. In the `fuzzy' DM region, CMB-S4 will allow for percent-level sensitivity to the axion mass fraction, improving significantly on current constraints. For {\sl Planck} data alone, neutrino degeneracies significantly degrade sensitivity to axions, even at the 1$\sigma$ level. In contrast, CMB-S4 constraints remain robust to varying neutrino mass in the `fuzzy' region. The solid and dashed lines show the $2\sigma$ and $1\sigma$ exclusion limits, i.e. the lowest axion fraction that could be excluded at those masses. \label{fig:axions}}
\end{figure}
Identifying dark matter (DM) remains one of the outstanding cosmological challenges of the current age.
While searches for direct or indirect evidence of a dark matter candidate continue \cite{Buckley:2013bha,Cushman:2013zza}, the effect of dark matter on cosmological observables provides a complementary approach to constraining the dark sector.

In the face of increasingly strict experimental limits to Weakly Interacting Massive Particle (WIMP) DM, axions are re-emerging as a popular alternative (see Ref.~\cite{Marsh:2015xka} for an extensive review of axions). Cosmological axion production can proceed through decays of exotic particles (e.g. moduli) or topological defects, thermal production from the standard-model plasma, or coherent oscillation around a misaligned (from the vacuum state) initial value, known as vacuum realignment. If axions are also produced because of non-vanishing matter couplings, a relativistic population can be produced, contributing to the relativistic energy density in the early universe (parameterized by a generic parameter $N_\mathrm{eff}$, describing the number of \textit{relativistic degrees of freedom}). Constraints on these axion models were presented in Refs. \cite{Acharya:2010zx,Weinberg:2013kea,Iliesiu:2013rqa,Baumann:2016wac}. 

Vacuum realignment is the only axion production mechanism that occurs independent of assumptions about axion couplings or inflationary physics, and produces an extremely cold population of axions, in contrast with other mechanisms. Here, we consider only axions produced by vacuum realignment.\footnote{We do vary $N_\mathrm{eff}-3.046$, but without bias as to its physical nature.} Ultralight axions (ULAs) produced via vacuum realignment with masses in the range $10^{-33}~\mathrm{eV}\leq m_{a}\leq 10^{-20}~\mathrm{eV}$ are well motivated by string theory, and can contribute to either the dark matter or dark energy components of the Universe, depending on their masses \cite{Marsh:2015xka}. 

They are distinguishable from standard dark energy (DE) and cold dark matter (CDM) using cosmological observables such as the cosmic microwave background (CMB) temperature and polarization power spectra, the matter power spectrum (as probed using the correlations of galaxy positions and shapes) and the weak gravitational lensing of the CMB. Constraints on the allowed contribution of ULAs to the total DM component using these observables provide a test of the CDM scenario.

A key goal of future cosmological experiments is to measure the sum of the neutrino masses, $\Sigma m_\nu$ (see Ref.~\cite{Lesgourgues:2012uu} for a review of neutrino cosmology). The current bound on $\Sigma m_\nu$ from ground-based oscillation experiments is $\Sigma m_\nu \gtrsim 0.06~\mathrm{eV}$ \cite{Otten:2008zz}. Current cosmological neutrino bounds indicate that $\Sigma m_\nu < 0.23~\mathrm{eV}$ at 95\% confidence, using data from Planck \cite{Ade:2015xua} and measurements of Baryon Acoustic Oscillations (BAO) from the Baryon Oscillation Spectroscopic Survey \cite[BOSS,][]{Beutler:2014yhv}.

Forecasted constraints for neutrino masses are that $\sigma(\Sigma m_\nu)= 15~\mathrm{meV}$ for a fiducial model with $\Sigma m_\nu = 60~\mathrm{meV}$, for a CMB-S4-like experiment and BAO measurements from a `DESI-like' survey \cite{Abazajian:2013oma}, promising a 4$\sigma$ detection of neutrino mass \cite{Allison:2015qca}. Much of this improvement is driven by weak gravitational lensing of the CMB, in particular at high multipoles $\ell\gtrsim 1000,$ although the change in the lensing convergence power is of order 25\% even at low multipoles. The lensing deflection power-spectrum is determined from $4$-pt functions of CMB maps, extracting a factor of  $\sim\sqrt{3}$ as much information from CMB experiments~\cite{Scott:2016fad}. 

The promise of CMB experiments in probing neutrino masses motivates us to wonder: will future CMB experiments offer dramatic improvements in sensitivity to axion parameters? Given the known similarity of ULA and massive neutrino imprints \cite{Marsh:2011bf} on cosmological observables at low mass ($m_{a}\lsim 10^{-29}~{\rm eV}$), how significant are ULA-neutrino degeneracies at CMB-S4 sensitivity levels and will they degrade our ability to do fundamental physics with the CMB? To answer these questions, we conduct a Fisher-matrix analysis to explore the sensitivity of future CMB experiments to ULA masses, densities, and $\Sigma m_{\nu}$. We find that CMB-S4 will allow a $2-5\sigma$ detection of axion mass fractions that agree with pure {\sl Planck} limits, covering an axion mass range of $10^{-32}~{\rm eV}\lsim m_{a}\lsim10^{-24}~{\rm eV}$. 

Near the top of this range, CMB-S4 will break the degeneracy of axions and CDM. Sensitivity persists (but tapers off) towards higher axion masses of $m_{a}\sim 10^{-23}~{\rm eV}$. CMB-S4 will push CMB tests of the ULA hypothesis towards the mass range probed by subtle observables, like the size of DM-halo cores and the number of missing Milky-Way satellites. In the ``dark-energy-like" (``DE-like" ULAs henceforth) ULA regime ($m_{a}\lsim 10^{-29}~{\rm eV}$) we find that the the ULA mass fraction is degraded by  degeneracies with the sum of the neutrino masses, but that this degeneracy disappears at higher masses. We find also that future measurements of the Hubble constant could break this degeneracy. We denote ULAs in the mass range $10^{-29}~{\rm eV}\lsim m_{a}\lsim 10^{-25}~{\rm eV}$ as ``fuzzy DM", and those with $m_{a}\gsim 10^{-25}~{\rm eV}$ as ``dark-matter-like" (or DM-like).

We find that measurements of the lensing-convergence power spectrum $C_{\ell}^{\kappa \kappa}$ drive much of the improvement in sensitivity; if lensing is omitted, the fractional error bar on the axion mass fraction degrades by a factor of $\sim 3-5$ in the `fuzzy' regime. Finally, we explore the dependence of our results on CMB-S4's experimental design parameters. 

We begin this paper by summarizing the physics and cosmology of ULAs and neutrinos in Section~\ref{sec:axion_cosmo}. In Section~\ref{observables}, we discuss the effects of ULAs and neutrinos on cosmological observables (e.g., the CMB's primary anisotropies and its lensing-deflection power spectrum), as well as the degeneracies between axions and cosmic neutrinos. Our assumptions about future data, forecasting techniques, and key science results are presented in Section \ref{sec:results}. We conclude in Section \ref{conclusion}.

All power spectra presented here were computed using the \textsc{AxionCAMB} code, a modification to the CMB anisotropy code \textsc{CAMB} \cite{cambnotes}, which is described in Appendix \ref{code_appendix}, is publicly available, and was used to obtain the ULA constraints of Ref.~\cite{Hlozek:2014lca}.\footnote{The code may be downloaded from \url{http://github.com/dgrin1/axionCAMB}.} In Appendix \ref{nonlinear_appendix}, we discuss the computation of the nonlinear matter power-spectrum (relevant for understanding the effect on weak lensing on the CMB).

\section{Review of Axion and Neutrino Cosmology \label{sec:axion_cosmo}}

This section provides a brief introduction to axion physics, as well as the cosmology of axions and neutrinos (reviewed in greater depth by Refs.~\cite{Marsh:2015xka,Hlozek:2014lca} and \cite{Lesgourgues:2012uu}, respectively).

In this work we model the axion as a scalar field $\phi$. The dynamics of the scalar field are set by its potential, which we assume for simplicity to be a $V(\phi) \simeq \frac{1}{2}m^2\phi^2$ potential. Hence the equation of motion for the homogeneous ULA is:
\begin{align}
\ddot{\phi}_0+2\mathcal{H}\dot{\phi}_0+m_a^2 a^2 \phi_0&=0,\label{homo_eom}
\end{align}
where the conformal Hubble parameter is $\mathcal{H}=\dot{a}/a=aH$, and dots denote derivatives with respect to conformal time. 
 
At early times the axion is slowly rolling and has an equation-of-state of $w_{a}\equiv P_{a}/\rho_{a}\simeq -1$. It therefore behaves like a cosmological constant, with roughly constant proper energy density as a function of time. $H$ decreases with the expansion of the universe and at a time $a_{\rm osc}$ such that $m_a\approx 3H(a_{\rm osc})$ the axion field begins to coherently oscillate about the potential minimum. 

The relic-density parameter $\Omega_{\rm a}$ is given by
\begin{equation}
\Omega_{\rm a}=\left[\frac{a^{-2}}{2}\dot{\phi}_0^2 + \frac{m_a^2}{2}\phi_0^2 \right]_{m_a=3H}a_{\rm osc}^{3}/\rho_{\rm crit},\label{homorelic}
\end{equation}
where $\rho_{\rm crit}$ is the cosmological critical density today. This production mode is known as the vacuum realignment, or misalignment, mechanism. 

In the early universe, neutrinos, like other weakly interacting particles, are coupled to the cosmological fluid until the weak interaction rate falls below the temperature of the universe, which is decreasing due to its expansion. This occurs at around $T\approx 1$ MeV. At this time, the neutrinos then decouple from the plasma. Massive neutrinos contribute to the energy density of the Universe as 
\be 
\Omega_\nu h^2 = \frac{\Sigma m_\nu }{93.14\mathrm{eV}}. \label{eqn:mass_neu}
\ee 
Massive neutrinos behave as radiation at early times (energy density scaling as $a^{-4}$). When the temperature drops below the neutrinos mass, they behave like matter (energy density scaling as $a^{-3}$). Thus, depending on the mass, massive neutrinos can change the time of matter-radiation equality, and alter the matter density at late times. Upper bounds on the mass of standard model neutrinos imply that they have a cosmologically non-negligible free-streaming length caused by their relativistic motion at early times. For wavenumbers $k>k_{\rm fs}$, neutrino clustering is suppressed relative to that of ordinary matter, leading to decreased structure formation for larger $\sum m_\nu$ (given a fixed late-time DM content).

ULAs also suppress structure formation at large wavenumbers, $k\gtrsim k_{\rm m}$, through their scale-dependent sound speed \cite{Marsh:2010wq,Hlozek:2014lca}:
\begin{eqnarray}
c_a^{2}=\left\{\begin{array}{ll}
\frac{k^2}{4m_a^2a^2}&\mbox{if $k\ll k_{m}\equiv 2m_{a}a$},\\
1&\mbox{if $k\gg k_{m}$}.\end{array}\right.\label{heuristic_cs}\end{eqnarray}
The wave number $k_{\rm m}$ is mass dependent, moving to large length scales as the axion mass decreases.  It is important to note that the axion suppression of structure and the suppression from neutrinos have very different physical origins: ULAs suppress structure growth below the Jeans length due to their wave-like nature, while neutrinos do so because of their large thermal velocities. 

In addition to the contribution of a massive neutrino species, we will investigate the degeneracies between vacuum-alignment ULAs and additional massless neutrinos and other ``dark radiation'' through the \textit{relativistic degrees of freedom} ($N_\mathrm{eff}$), parameterized relative to the photon energy density, $\rho_\gamma$, as:
\be
\rho = N_\mathrm{eff}\frac{7}{8}\left(\frac{4}{11}\right)^{4/3}\rho_\gamma.
\ee
For useful descriptions of the physics of $N_{\rm eff}$ on the CMB, see Refs.~\cite{2013PhRvD..87h3008H,2015PhRvL.115i1301F}.

As noted above, ULAs produced by vacuum realignment do \emph{not} contribute to $N_\mathrm{eff}$. Axions produced by other mechanisms, however (such as thermal freeze-out or heavy particle-decay) constitute a separate population of relativistic axions, and \emph{do} contribute to $N_\mathrm{eff}$ \cite{Acharya:2010zx,Weinberg:2013kea,Iliesiu:2013rqa,Baumann:2016wac}. It is important to note that $N_\mathrm{eff}$ does not distinguish between fermions and bosons (although other cosmological observables could. See, for example Ref.  \cite{Hannestad:2005bt}), nor on the production mechanism of the additional radiation. Thus additional relativistic neutrinos and axions are completely degenerate in cosmological terms: because of this we consider varying $N_\mathrm{eff}$ completely generically.

The lightest  vacuum-realignment ULAs ($m_a < 10^{-30}~\mathrm{eV}$) are degenerate with a DE-like component in the universe, and generate a late-time integrated Sachs-Wolfe (ISW) \cite{Sachs:1967er} effect in the CMB \cite{Crittenden:1995ak,Coble:1996te,Hlozek:2014lca,Boughn:2004zm}. They also change the background expansion rate of the universe, altering the angular diameter distance to the last-scattering surface. This affects the position of the peak in a similar manner to how $N_\mathrm{eff}$ alters the position of the peak. Hence we expect a partial degeneracy between ULAs and $N_\mathrm{eff}$ for the lightest ULAs.

\section{CMB observables \label{observables}}

The main effects of ULAs in the temperature power in the multipole range relevant to {\sl Planck}, and in the linear galaxy power spectrum, were discussed in detail in Ref.~\cite{Hlozek:2014lca}. Primary CMB power spectra, matter power-spectra, and lensing convergence power-spectra for ULAs are all computed using the \textsc{AxionCAMB} code, which was used to obtain the results of Ref.~\cite{Hlozek:2014lca} and is described in the Appendix~\ref{code_appendix} of this paper.

\subsection{The CMB-damping tail, distance measures, and neutrino degeneracies}

In order to interpret forecasts on the allowed values of the energy density in ULAs and the degeneracies with neutrinos, we highlight the similarities and differences between the two components at the level of effects on the cosmological observables. This comparison was made for galaxy surveys in Ref.~\cite{Marsh:2011bf}, and was also discussed in Ref.~\cite{Amendola:2005ad}. 

\begin{figure*}[htbp!]
\begin{center}
$\begin{array}{ll}
\includegraphics[width=\columnwidth,trim = 0mm 0mm 0mm 60mm, clip]{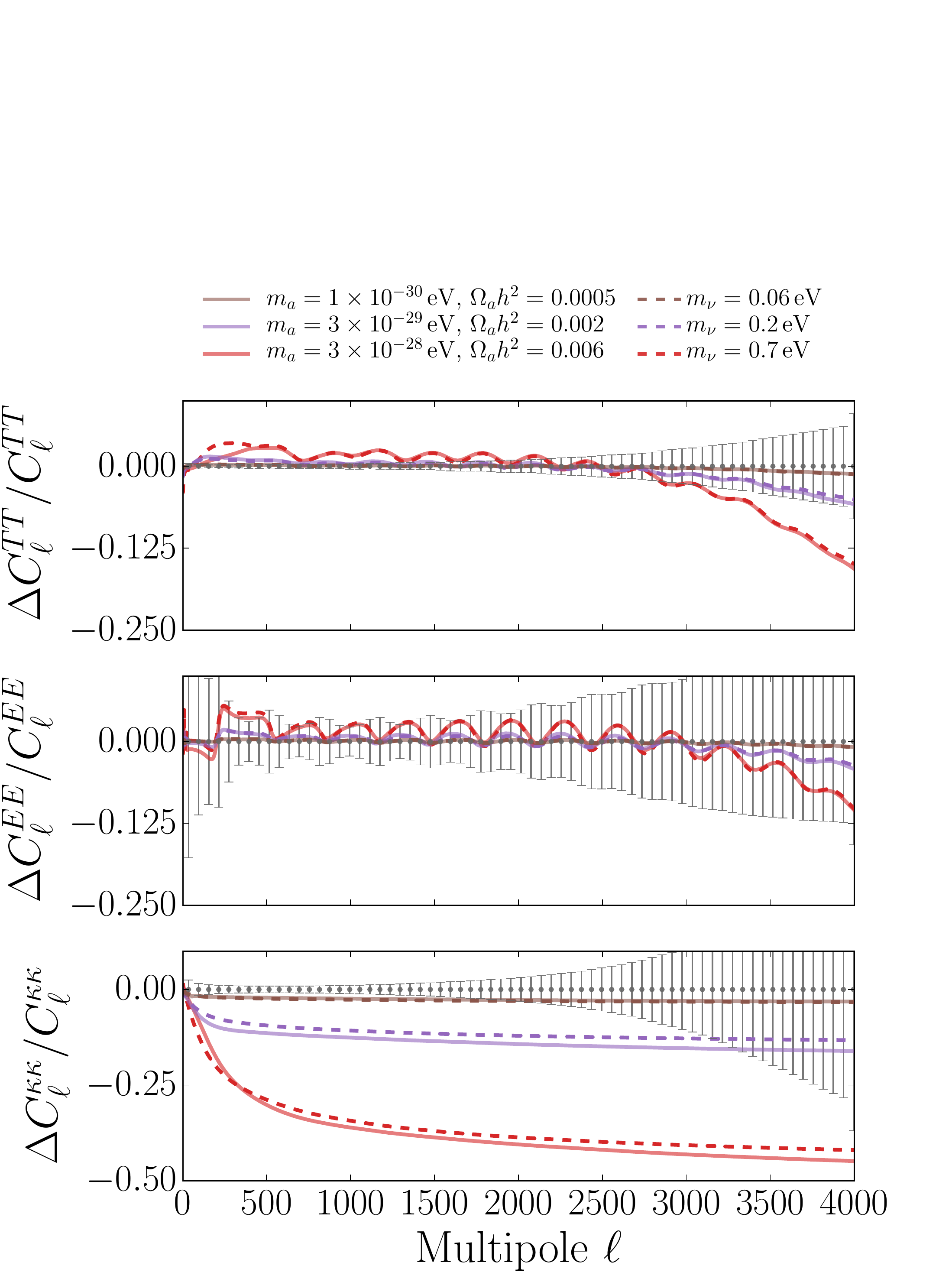}&
\includegraphics[width=\columnwidth,trim = 0mm 0mm 0mm 60mm, clip]{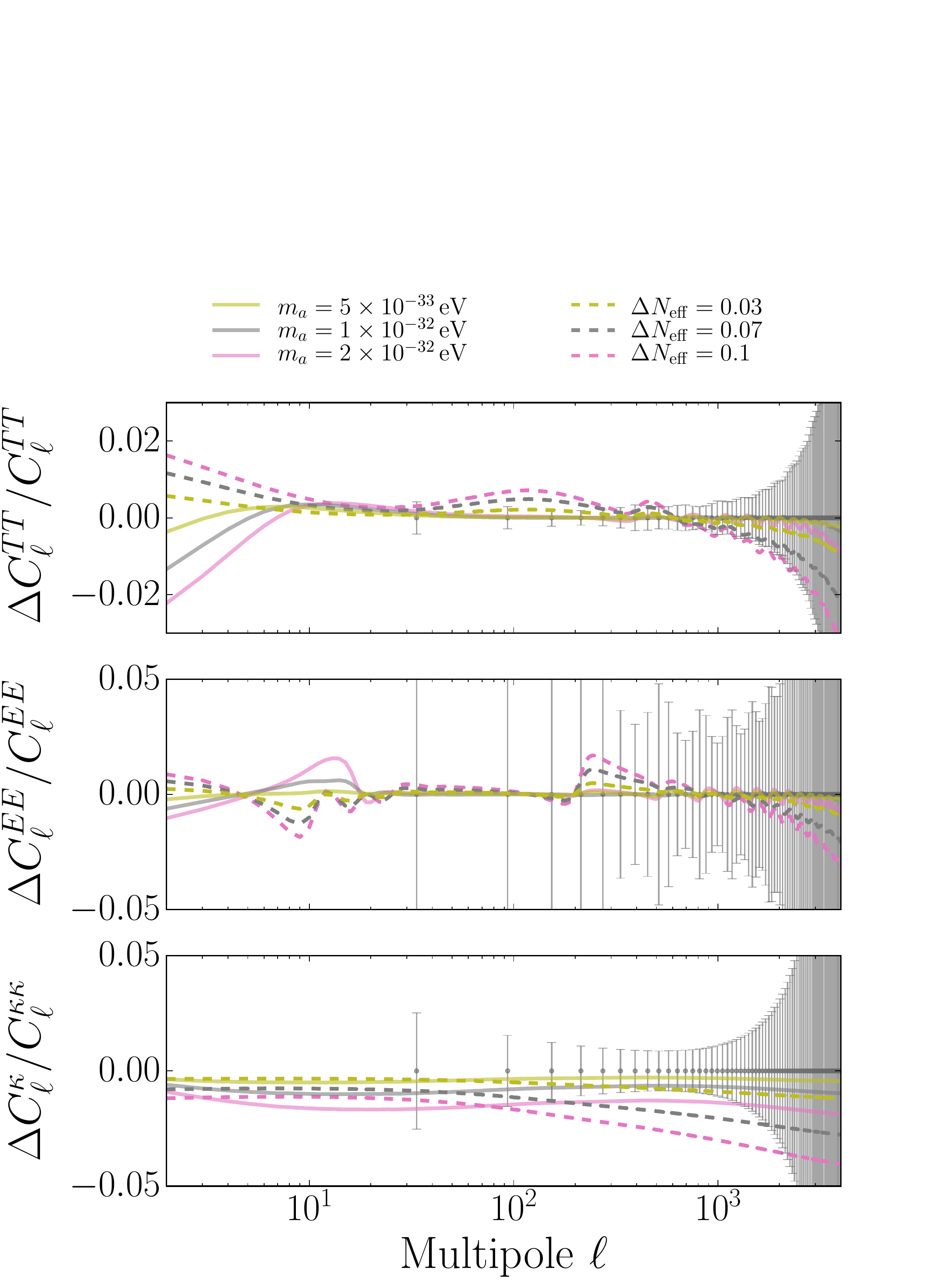}
\end{array}$
 \caption{\textbf{Relative differences between axion effects and other cosmological parameters:} The error bars shown are for a `CMB-S4-like' survey as described in Table~\ref{table:net}. \textit{Left:} Comparing DM-like ULAs to massive neutrinos, holding the total matter density and sound horizon fixed. For each neutrino model, there is a corresponding ULA model that produces almost degenerate effects in all observables. Both massive neutrinos and ULAs produce the largest effects in the lensing convergence power, where effects of order $\Delta C^{\kappa\kappa}_\ell/C^{\kappa\kappa}_\ell \simeq 25\%$ occur even at low multipoles. \textit{Right:} Comparing DE-like ULAs (with $\log_{10}(m_a/\mathrm{[eV]}) < -30$) to additional massless neutrinos, holding only the sound horizon fixed. The ULA energy density is $\Omega_a h^2=0.002$. These types of models do not display any significant degeneracies. Note that the $\ell$-axis of the right panel is shown in log scale, while the left panel is linear.  \label{fig:observables_massive}}
\end{center}
 \end{figure*}

ULAs and neutrinos affect the expansion rate, changing the angular size of the sound horizon, $\theta_s$, at fixed Hubble constant, $h$. Consider the case of one additional massive neutrino eigenstate with $\Sigma m_\nu=0.06~\mathrm{eV}$ and $N_\mathrm{massive}= 1$, $N_\mathrm{massless} = 2.046$. This neutrino is relativistic throughout the radiation era, but behaves like matter at late times. The main effect of this on the high-$\ell$ acoustic peaks is to increase the angular size of the sound horizon. This can be compensated by \textit{reducing} the Hubble constant from $h=0.6715$ in  to $h=0.6685$, in order to hold $\theta_s$ fixed (relative to a $\Sigma m_\nu=0$ model). ULAs also change the expansion rate relative to pure CDM due to the early $w_a=-1$ behaviour: holding $\theta_s$ fixed requires a reduction in $h$ just as for neutrinos~\cite{Marsh:2011bf,Hlozek:2014lca}. 

In Figure~\ref{fig:observables_massive} we show the relative difference in CMB auto power spectra for temperature, T, E-mode polarization, and lensing convergence, $\kappa$, for ULA and neutrino models compared to a reference $\Lambda$CDM model:
\be
\frac{\Delta C_\ell}{C_\ell} = \frac{(C_{\ell}^{\rm model}-C_\ell^{\rm ref.})}{C_\ell^{\rm ref.}} \, .
\ee
The reference model contains $N_{\rm eff}=3.046$ massless neutrinos, and no ULAs. Massive neutrinos are introduced as a single massive eigenstate, i.e. $N_{\rm massive}=1$, $N_{\rm massless}=2.046$, with the energy density today fixed by the mass in as in Eq.~(\ref{eqn:mass_neu}). ULAs are introduced with a free mass and energy density, and are chosen to mimic as closely as possible the neutrino models in the observables. 

The ULA and neutrino models are chosen to keep the total matter density, $\Omega_m h^2=\Omega_c h^2+\Omega_b h^2+\Omega_a h^2+\Omega_\nu h^2$, and sound horizon, $\theta_s$, fixed. Under these conditions, the effects of ULAs and massive neutrinos on the CMB observables are remarkably similar, and it is clear  that there are parts of parameter space where significant degeneracies exist. Were one also to vary the number of massive neutrinos, $N_{\rm massive}$, even more degeneracies would open up~\cite{Marsh:2011bf}. 

For example, we observe that a ULA model with $m_a=10^{-30}\text{ eV}$ and $\Omega_a h^2=0.0005$ is degenerate with the standard fiducial neutrino model with $m_\nu=0.06\rm{ eV}$. This ULA energy density occurs naturally (i.e. the axion misalignment angle $\theta \ll 1$) for $f_a\approx 3\times 10^{-2}M_{pl}\approx 7\times 10^{16}\text{ GeV}$: GUT-scale ULAs can be constrained by the CMB, but also have significant degeneracies with other cosmological components.

In the most massive neutrino model shown in Figure~\ref{fig:observables_massive}, $\Sigma m_\nu=0.7\text{ eV}$, holding the sound horizon fixed requires decreasing the Hubble constant to $h=0.6415$, while the corresponding axion model only requires $h=0.6635$. For the other reference models with lighter neutrinos, the change in $h$ required for ULAs and neutrinos is the same. Thus, in the case of relatively heavy ULAs and neutrinos, a local measure of $H_0$ can help break degeneracies. 

ULAs and massive neutrinos can produce $\mathcal{O}(10\%)$ effects in the temperature power at $\ell\gtrsim 3000$. This comes from the lensing-induced temperature power, which at high $\ell$ is approximately~\cite{Lewis:2006fu}: 
\begin{equation}
C_{\ell}^{\rm TT}\approx \ell^2 C_\ell^{\phi\phi}\int \frac{d\ell'}{\ell'}\frac{\ell'^4}{4\pi}\tilde{C}_{\ell'}^{\rm TT}\, ,
\end{equation}
where $\tilde{C}_\ell$ is the unlensed power, and $C_{\ell}^{\phi\phi}$ is the power spectrum of the lensing potential. 

The lensed temperature power in ULA and massive neutrino cosmologies is reduced compared to pure CDM by the suppression of clustering (free streaming for neutrinos, the Jeans instability for ULAs) and consequent reduction of the lensing contribution to $C_\ell^{\rm TT}$. This effect is likely of little importance observationally, as temperature power at such high multipoles becomes dominated by other secondaries, such as galactic foregrounds, and the Sunyaev-Zel'dovich effect, making the direct lensing contribution hard to measure. A similar effect is also seen in the E-mode polarization, which suffers less from foregrounds at high multipoles. The effects of massive neutrinos and ULAs on the lensed E-mode power at high-$\ell$ are relatively small, however, compared to the forecasted CMB-S4 error bars.

Both massive neutrinos and ULAs produce the largest effects at relatively low multipoles in the lensing convergence power, and this offers a very powerful observable to constrain the properties of DM beyond CDM. The lensing convergence power spectrum, $C_\ell^{\kappa\kappa}$, is a direct measurement of the DM distribution, and its scale dependence at high-$\ell$ measures the clustering properties of sub-dominant components of the DM. In Ref.~\cite{Allison:2015qca}, it was shown that the lensing convergence power drives the ability of future CMB experiments to measure the sum of neutrino masses. Figure~\ref{fig:observables_massive} shows that the lensing convergence power also provides a powerful method to constrain other departures from CDM, and measures the composition and clustering properties of DM over a wide range of scales. We will quantify this in detail in Section~\ref{constraints}, showing the gains in sensitivity given by CMB-S4 over {\sl Planck}, and how much of this gain is driven by lensing.
 
 Now consider the effect of additional massless neutrinos, parameterized by $\Delta N_{\rm eff}$, and DE-like ULAs (i.e. those for which $w_a=-1$ for some period during the matter dominated era). The effects of these models on CMB observables are also shown in Figure~\ref{fig:observables_massive}. We notice the well-known effect that $\Delta N_{\rm eff}\neq 0$ increases the amount of damping in the CMB at high-$\ell$. Since we include radiation in the closure budget, there is also reduced overall matter power, and consequently reduced lensing power. DE-like ULAs affect the lensing largely through the expansion rate and scale-dependence of the growth at low-$z$. This has a knock-on effect of slightly reducing TT and EE power at large $\ell\gtrsim 1000$ from reduced lensing, and in some cases creates a partial degeneracy with $N_{\rm eff}$ on these scales.

There are $\mathcal{O}(1\%)$ effects in the ${\rm EE}$ power for $N_{\rm eff}$ and DE-like ULAs at $\ell\approx 10$, the ``reionization bump'', caused by the different expansion histories and matter budgets in these models. The low-$\ell$ effects of $\Delta N_{\rm eff}$ and DE-like ULAs in TT and EE are opposite in sense, which predicts the degeneracy direction if such multipoles are included - here combining temperature and polarization data helps break the degeneracy. We also notice $\mathcal{O}(1\%)$ effects of $N_{\rm eff}$ at $\ell\approx 100$ in $EE$ at the ``recombination bump'', similarly caused by effects on the expansion rate. DE-like ULAs do not affect recombination relative to $\Lambda$CDM, since they behave entirely like the cosmological constant $\Lambda$ at this epoch by definition.

For $\Delta N_{\rm eff}\neq 0$ and DE-like ULAs, we have adjusted $H_0$ to hold the sound-horizon fixed. This serves to further physically distinguish the models. Massless neutrinos decrease $\theta_s$ and require an increase in $H_0$ to hold it fixed: hence a preference for $\Delta N_{\rm eff}\neq 0$ is sometimes found to reconcile CMB (lower) and other (higher) measures of $H_0$ \citep[e.g.][]{MacCrann:2014wfa}. On the other hand, we introduce DE-like ULAs with constant $\Omega_c h^2$, and as such they come out of the DE budget. As described in detail in Ref.~\cite{Hlozek:2014lca}, they require reduced $H_0$ to hold $\theta_s$ fixed, and lead to a non-$\Lambda$ effect on the late-time ISW effect at low $\ell$. In the most extreme cases shown, $\Delta N_{\rm eff}=0.1$, $m_a=2\times 10^{-32}\text{ eV}$ the change in $h=\pm 0.1$ respectively. Accurate local measures of $H_0$ can improve constraints on DE-like ULAs substantially \citep[e.g.][]{Freedman:2010xv, Riess:2016jrr}, but high-$\ell$ CMB experiments such as CMB-S4 will add little to constraints on them compared to {\sl Planck}. We discuss quantitatively the inclusion of a prior on $H_0$, in addition to CMB-S4, in Section~\ref{constraints}.

In conclusion on this topic, we do not expect significant degeneracies between additional massless neutrinos and DE-like ULAs, while we expect significant degeneracies between heavier ULAs and massive neutrinos. Via lensing, CMB-S4 should allow detection of neutrino mass, and greatly improve constraints on intermediate mass ULAs. CMB-S4 should also substantially improve constraints on $\Delta N_{\rm eff}$ by more precise measures of the damping tails. Including $H_{0}$ measurements should improve limits on DE-like ULAs, and break remaining degeneracies.
 
\subsection{Lensing deflection power and non-linear clustering }
\label{sec:nonlinear}

The largest deviation from standard $\Lambda$CDM caused by ULAs in the lensing deflection power occurs on small scales. Here one must take some care as both non-Gaussian noise in the experimental setup, and the theoretical modeling of nonlinear lensing add a systematic error to any inferred constraints on DM properties. This problem is particularly acute for more massive ULAs ($m_a\gtrsim 10^{-25}\text{ eV}$), which undergo non-linear clustering on observationally relevant scales or redshifts and can contribute a large fraction to the total DM abundance.

The lensing deflection power, $C_\ell^{\kappa\kappa}$, depends on the integral along the line-of-sight of the Newtonian potential power spectrum, $\mathcal{P}_{\Psi}(k,z)$~\cite{Lewis:2006fu}. These non-linear clustering contributions such that non-linear effects before important on larger angular scales in $C_\ell^{\kappa \kappa}$ than they do for $C_{\ell}^{\rm TT}$.

The lensing power on all multipoles is dominated by effects at $z\lesssim 10$. For multipoles $\ell\approx 1000$ the integral kernel peaks at $z\approx 2$. In terms of wavenumber, $k$, multipoles $\ell\gtrsim 1000$ are dominated by contributions from, $k\gtrsim 0.1\text{ Mpc}^{-1}$, where density perturbations are becoming non-linear. On these sub-horizon scales, the power spectrum of the Newtonian potential is determined from the matter power spectrum via Poisson's equation. Non-linearities in the matter clustering in this range of redshifts and wavenumber lead to $\mathcal{O}(10\%)$ effects in the lensing power for $\ell\gtrsim 1000$.

The non-linear gravitational potential power spectrum (needed to compute $C_{\ell}^{\kappa\kappa}$ including nonlinear effects) is computed in \textsc{camb} using the expression (see Ref. \cite{2010GReGr..42.2197H} and references therein):
\begin{align}
\mathcal{P}_{\Psi,{\rm non-lin}}(k,z) &= \frac{P_{m,{\rm non-lin}}(k,z)}{P_{m,{\rm lin}(k,z)}}\mathcal{P}_{\Psi,{\rm lin}}(k,z) \\
&\equiv \mathcal{R}_{\rm nl}(k,z)\mathcal{P}_{\Psi,{\rm lin}}(k,z)\, ,\label{eqn:psi_non-lin}
\end{align}
where $P_m(k,z)$ is the matter power spectrum, and non-linearities are computed using \textsc{halofit}~\cite{Smith:2002dz}, a code based on a fitting function, which is calibrated to N-body simulations of CDM (with Ref. \cite{Bird:2011rb} including massive neutrinos). One must therefore take extra care when exploring constraints on non-standard models from high-multipole lensing.\footnote{This does not just apply to non-standard DM models, such as ULAs. \textsc{halofit} is calibrated using power law initial conditions, and so care must also be taken  for models with features in the primordial power spectrum at high-$k$.} We discuss the non-linear modeling of the power spectrum further in Appendix~\ref{nonlinear_appendix}.

\begin{figure}[htbp!]
\begin{center}
\includegraphics[width=1.05\columnwidth]{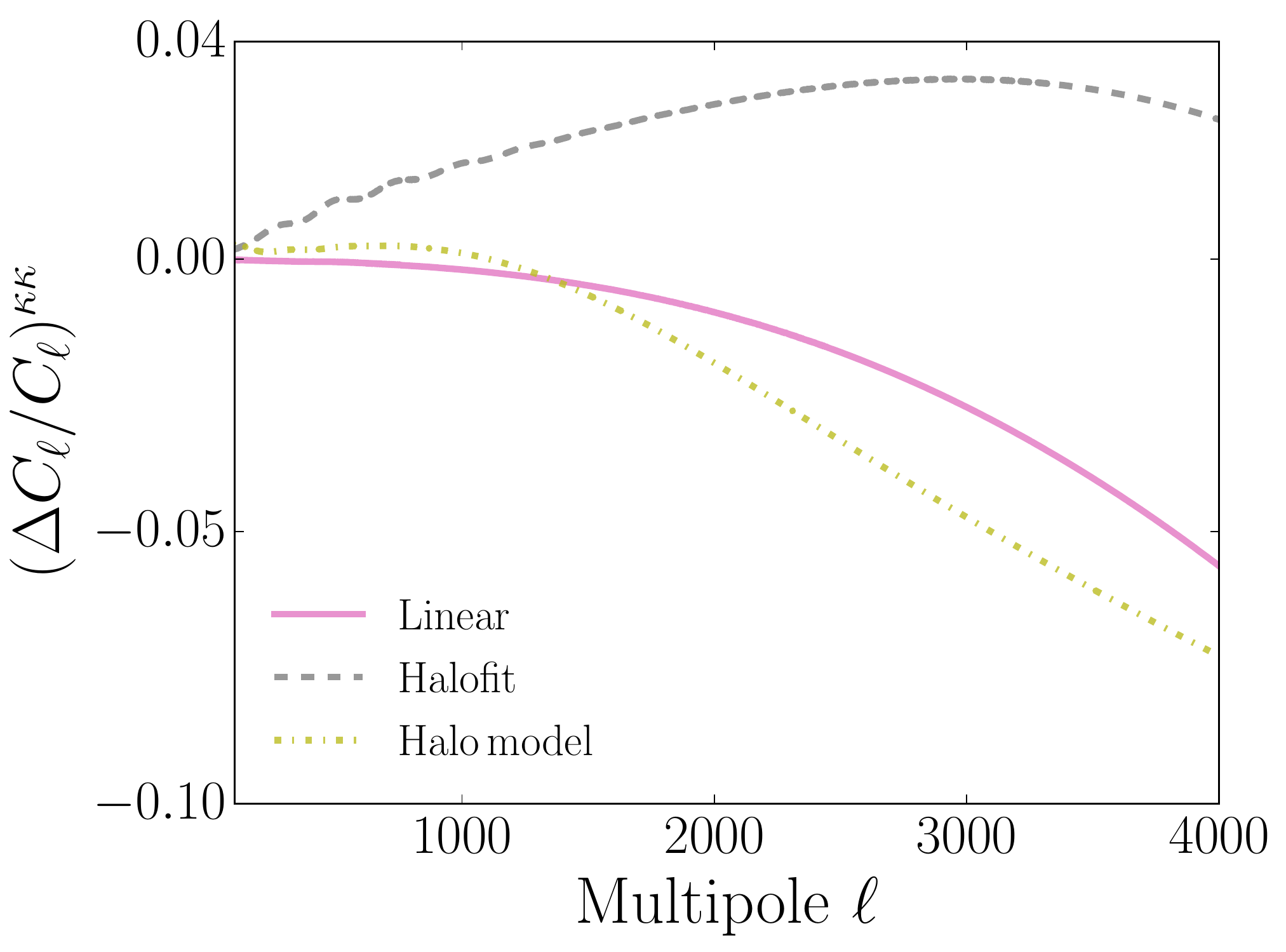}
 \caption{\textbf{Comparison of ULAs to CDM in lensing deflection power for different models of non-linearities, where ULAs with $\mathbf{m_a=10^{-23}}\text{ eV}$ constitute all the DM.} The unphysical power increase in the \textsc{halofit} power for ULAs, seen in Figure~\ref{fig:tk_halofit_halomodel}, causes a similar unphysical increase in lensing power compared to the halo model. On the other hand, linear theory captures the sign and approximate magnitude of the effect seen in the halo model. Thus when forecasting constraints at high ULA mass we choose to use linear theory lensing as a reasonable approximation for the Fisher matrix derivative. \label{fig:lensing_halofit_halomodel}}
\end{center}
 \end{figure}

We now assess how the non-linear modeling affects the lensing deflection power of ULAs.  Figure~\ref{fig:lensing_halofit_halomodel} shows the lensing power ratio $(\Delta C_\ell/C_\ell)^{\kappa\kappa}$ for $m_a=10^{-23}\text{ eV}$ assuming that either ULAs or CDM (but not both) constitute all of the DM. We compare linear theory, \textsc{halofit}, and the halo model for ULAs of Ref.~\cite{Marsh:2016vgj}. For illustration, we consider the lensing deflection power from the halo model under the Limber approximation (which is accurate for high-$\ell$ where non-linearities become important) \cite{2010GReGr..42.2197H}:
\begin{equation}
C_{\ell}^{\phi\phi} = \frac{8\pi^2}{\ell^3}\int_0^{z_{\rm rec}}dz \mathcal{P}_{\Psi}(\ell/x,z) x \frac{dx}{dz}\left(\frac{x_{\rm rec}-x}{x_{\rm rec}x} \right)^2 \,.
\end{equation}
where $x=x(z)$ is the comoving distance to redshift $z$.\footnote{We emphasize that the halo model for ULAs is \emph{not} yet incorporated into \textsc{axionCAMB}. The halo model for CDM has only recently been incorporated into \textsc{camb}. Our comparisons here attempt to use the halo model to motivate approximations appropriate to forecasting. Proper inclusion of non-linearities in real data analysis of CMB-S4 will be crucial to avoid bias caused by incorrect modeling.}

\begin{figure*}[htbp!]
\includegraphics[width=1.8\columnwidth]{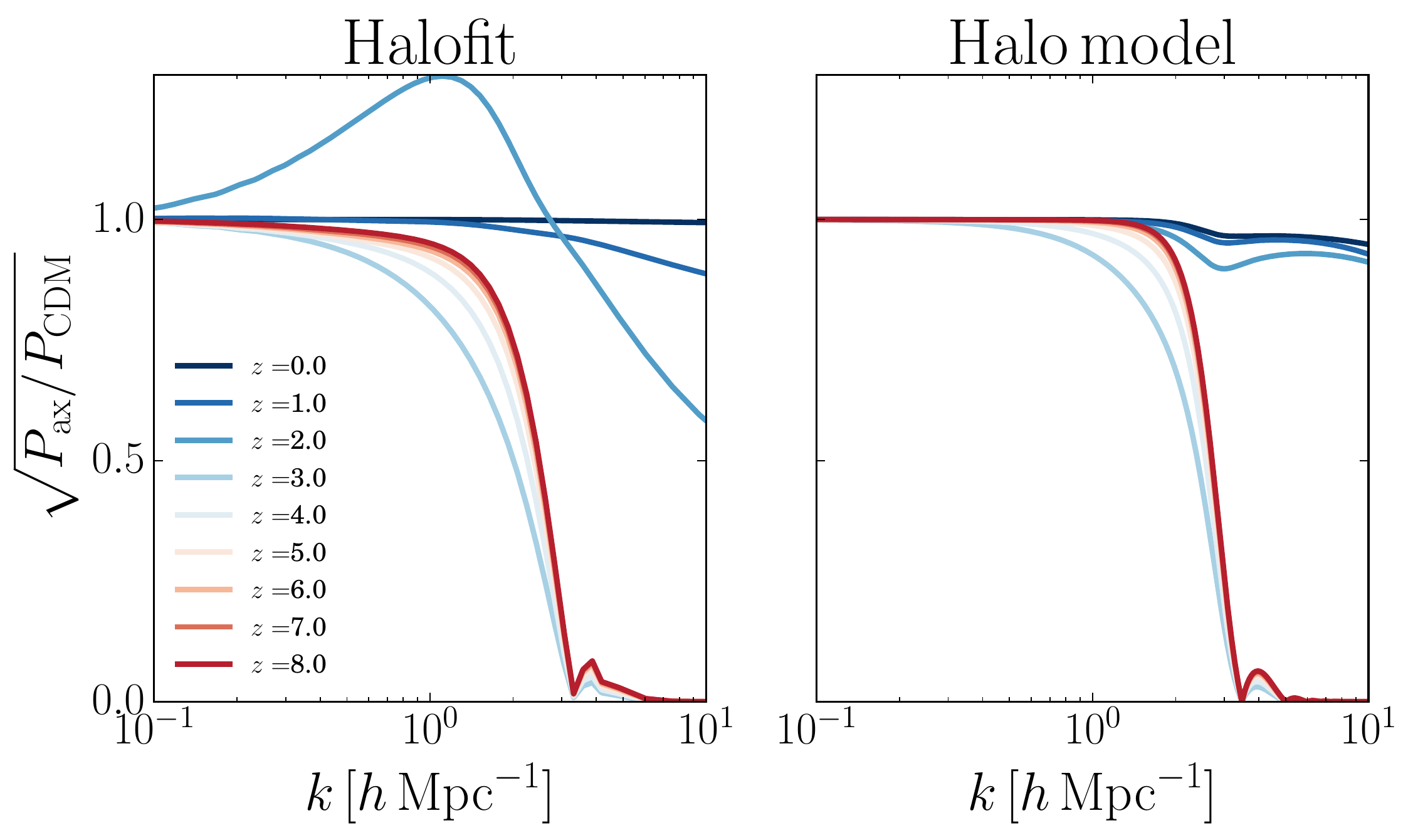}
 \caption{\textbf{Comparison of power spectrum ratios for \textsc{halofit} and the halo model of Ref.~\cite{Marsh:2016vgj}\footnote{Available online at: https://github.com/DoddyPhysics/HMcode.} for $m_a=10^{-23}\text{ eV}$ and ULAs as all the DM.} The halo model is cut to set the power to linear if $\sigma^2<1$ to make a fair comparison. Non-liner clustering begins at $z=2$. \textsc{halofit} applied to non-CDM models gives an unphysical boost in power at the onset of non-linearities, which is passed on to the lensing power, Figure~\ref{fig:lensing_halofit_halomodel}. Differences between the halo model and \textsc{halofit} at high $z$ are due to the quantitiative differences between the \textsc{axionCAMB} transfer function and the combination of Refs.~\cite{Eisenstein:1997ik,Hu:2000ke} analytic fits used in the halo model. \label{fig:tk_halofit_halomodel}}
\end{figure*}

Figure~\ref{fig:tk_halofit_halomodel} shows the overdensity ratio of ULAs to CDM, $\sqrt{P_{\rm ULA}(k,z)/P_{\rm CDM}(k,z)},$ over a range of scales and redshifts for a pure ULA DM model with $m_a=10^{-23}\text{ eV}$. In this model, perturbations in the axion energy density go non-linear for $z<3$, where non-linear collapse reduces the power suppression relative to CDM for $k\gtrsim 1\, h\text{ Mpc}^{-1}$. 

We notice that \textsc{halofit} introduces a large feature, increasing the power at the non-linear scale, $(k_{\rm nl},z_{\rm nl})$. Such a feature is not seen in the halo model, and is thus suspected to be an unphysical artifact introduced purely by the fitting functions of \textsc{halofit} -- calibrated to CDM and not a good description of ULAs at this scale. This unphysical boost in the matter power caused by \textsc{halofit} leads to a similarly unphysical increase in the lensing deflection power in Figure~\ref{fig:lensing_halofit_halomodel}. The effect seen in the (presumably more correct) halo model is that ULAs always decrease the lensing deflection power relative to CDM. Furthermore, perhaps surprisingly, the sign and approximate magnitude of the \emph{relative} effect of ULAs compared to CDM on the lensing deflection power in the full halo model is well captured by linear theory.

The above observation - that linear theory captures the relative effects of high mass ULAs on weak lensing better than \textsc{halofit} - determines how we decide to treat non-linear modeling in our forecasts (see also Appendix~\ref{nonlinear_appendix}). We choose by default to \emph{perform all forecasts with non-linear lensing turned off}. This choice is expected to give the right sign and approximate magnitude for Fisher-matrix derivatives for high mass ULAs, while non-linear modeling is not expected to be important at low mass, where ULAs do not non-linearly cluster on the relevant redshift range. 

\section{Results}\label{sec:results}
This section contains our assumptions, methodology, and key science results. In Sec.~\ref{data}, we lay out the assumptions made about CMB-S4 and Fisher-matrix techniques used to obtain our results. In Sec.~\ref{constraints}, we present our conclusions about the sensitivity of CMB-S4 to ULAs, the improvement over {\sl Planck}, the role of CMB weak lensing in driving sensitivity improvements, and explore degeneracies with neutrinos. Finally in Sec.~\ref{survey_optimize}, we explore how varying potential CMB-S4 survey parameters (sky coverage, noise level, and beam width) affects the conclusions of Sec.~\ref{constraints}. 

\subsection{Data and Surveys\label{data}}
The current best constraints on the axion fraction comes from a combination of the primary CMB ({\sl Planck}, SPT, and ACT TT power spectra, as well as low-$\ell$ {\sl WMAP} polarization data) with the WiggleZ galaxy redshift survey \cite{Hlozek:2014lca}. 

We consider future constraints from a `CMB-S4-like' survey as discussed in the recent Snowmass proposal \cite{Abazajian:2013oma}, with observational parameters specified in Table~\ref{table:net}. The exact specification of a CMB-S4
experiment is still under development. Provided it covers a significant fraction of the sky with reasonable noise levels, CMB-S4 promises to be an incredible instrument with which to test the dark sector.
In Section~\ref{survey_optimize} we test for the dependence of the constraints on the survey parameters. 

We forecast assuming a fiducial set of cosmological parameters:
\begin{equation}
\mathbf{\Xi} = \left\{\Omega_bh^2, \Omega_dh^2, H_0, A_s, n_s, \tau, m_a \Omega_a/\Omega_d \right\},
\end{equation}
where $\Omega_bh^2$ parameterizes the physical baryon density of the universe, $\Omega_dh^2$ is the energy density of the dark sector including axions, $H_0$ is the Hubble parameters in units of ${\rm km\, s}^{-1}{\rm Mpc}^{-1}$,  $A_s, n_s$ are the amplitude and spectral index of the scalar density fluctuations and $\tau$ is the optical depth to decoupling. As described above, the fraction of the dark sector made of axions (at a specified fixed axion mass $m_a$ in units of ${\rm eV}$) is given by $\Omega_a/\Omega_d$. The fiducial values and step sizes used for this model are shown in Table~\ref{tab:fiducial}.

\begin{table}[t!]
\begin{center}
\begin{tabular}{ccc}
\hline
\hline
Parameter&Fiducial value& Step size\\
\hline
$\Omega_bh^2$ & 0.02222 & 0.0001\\
$\Omega_dh^2$ &0.1197 & 0.001  \\
$H_0\,[{\rm km\, s}^{-1}{\rm Mpc}^{-1}]$ & 69.0& 0.1\\
$A_s $ &$ 2.1955\times 10^{-9}$&  $2.0\times 10^{-11}$\\
$n_s$ & 0.9655& 0.005 \\
$\tau$ & 0.06 & 0.01 \\
$m_a\,[\mathrm{eV}]$ & $ 10^{-32} < m_a< 10^{-22}$ & [fixed per run]\\
$\Omega_a/\Omega_d $ & 0.02 & 0.005 \\
\hline
\end{tabular}
\caption{\textbf{Fiducial model and Fisher Matrix step sizes:} The base model considered and the step sizes used to compute the Fisher derivatives. The above model was also supplemented in parts by including the additional extensions of the parameters $\Sigma m_\nu=60\,\mathrm{meV}$ and $N_\mathrm{eff}=3.046$ which were varied with step sizes of $20\,\mathrm{ meV}$ and $0.05$ respectively. \label{tab:fiducial}}
\end{center}
\end{table}

Where necessary we include $\Sigma m_\nu\,\mathrm{[eV]}$ and $N_\mathrm{eff}$ as additional parameters in the model space.

We use Fisher-matrix techniques to forecast constraints on the parameters of interest \cite{Eisenstein:1998hr,Taylor:1997ag,Matsubara:2004fr,Bassett:2009uv}. The Fisher matrix translates uncertainties on observed quantities such as the lensing deflection or the CMB power spectrum into constraints on parameters of interest in the underlying model. 
The Fisher matrix is the expectation value of the second derivatives of the logarithm of the data likelihood with respect to the parameters $\mathbf{\Xi}:$

\begin{equation}
\mathscr{F}_{ij} = -\left\langle \frac{\partial^2 \ln P(\mathbf{D}|\mathbf{\Xi})}{\partial\Xi_i\partial\Xi_j}\right\rangle,
\end{equation}
where $\mathbf{D}$ is the data vector of either CMB measurements or lensing deflection, for example. 

For independent experiments (or if one has prior knowledge of the uncertainties on a parameter from a separate experiment) one can add individual Fisher matrices together to get a final Fisher matrix. In order to obtain 1- or 2-dimensional constraints on parameters (i.e. 1-D likelihoods or 2-D error ellipses), one marginalizes over the other nuisance parameters in the larger parameter space under consideration.

The Fisher matrix code ({\sc OxFish}) used to forecast the full set of observables including the lensing deflection is described in Ref.~\cite{Allison:2015qca}, modified to include the axion parameters, as described in Ref.~\cite{Hlozek:2014lca}.

We compared a five-point numerical derivative, \begin{eqnarray}
f'(x) &=& \left[8f(x+h)-8f(x-h) \right. \nonumber \\
&-& \left.  f(x+2h) + f(x-2h)\right]/12h,
\end{eqnarray} to the standard two-sided finite-difference derivative method and checked that the resulting parameter uncertainties were stable to the choice of derivative method. In addition, we demanded that the derivatives of the axion fraction converged to 0.1\% precision to set the step size used for finite-difference calculations.

We forecast the combination of our `CMB-S4-like' survey with{\sl Planck} temperature and polarization spectra that match the current sensitivities between the multipoles of $30< \ell < 2500$. This also allows us to assess the gains possible when moving from {\sl Planck} to {\sl Planck}+S4: Fisher-matrix forecasts are often somewhat more optimistic than sensitivities obtained in real experiments, and so we use Fisher forecasts for both {\sl Planck} and {\sl Planck}+S4 in order to conduct a fair comparison. For CMB-S4 we assume measurements of the TT,EE,TE primordial CMB spectra with an $\ell_\mathrm{min}=30$ and an $\ell_\mathrm{max}=4000$ for the EE, TE spectra and $\ell_\mathrm{max}=3000$ for the TT spectra. We include the lensing deflection power spectrum from both surveys between $30 < \ell < 3000$. 
For the low-$\ell$ data we use {\sl Planck} HFI `lowP' specifications, with slightly modified noise levels to ensure a prior on the optical depth of $\tau=0.01$.

We assume that the noise has a white power-spectrum, using the standard treatment \cite{Knox:1995dq}: 

\begin{equation}
N^{\alpha\alpha}_\ell = (\Delta \alpha)^2\exp{\left(\frac{\ell(\ell+1)\theta_\mathrm{FWHM}^2}{8\ln 2}\right)},
\end{equation} where $\alpha = {\rm T}$ or ${\rm E}$, labels the field of interest. $\theta_\mathrm{FWHM}$ is the beam full width half maximum, and the lensing deflection noise is estimated assuming a minimum-variance quadratic estimate of the lensing field as described in Ref.~\cite{Allison:2015qca}. We assume that relevant foregrounds have been removed on all scales up to $\ell = \ell_\mathrm{max}$. We don't include information from the BB lensing power-spectrum, as the assumption of nearly Gaussian fields (required for the validity of the Fisher-matrix formalism) breaks down for B-modes from lensing, which are produced by a scalar modulation of primordial E-modes, and is thus a higher order (and non-Gaussian) effect.

\begin{table}[t!]
\begin{center}
\begin{tabular}{cccc}
\hline
\hline
$f_\mathrm{sky}$& Beam size& $\Delta T$ & $\Delta E,B$ \\
& (arcmin)& ($\mu$K-arcmin)&($\mu$K-arcmin) \\
\hline
0.4& 1& 1&1.4\\ 
\hline
\end{tabular}
\caption{\textbf{Survey parameters considered for axion forecasts:} Survey sensitivity, assumed beam size and sky fraction for a possible `CMB-S4-like' survey. We test the dependence of the axion constraints on these parameters in Section~\ref{survey_optimize}.\label{table:net}}
\end{center}
\end{table}

\subsection{Forecasted sensitivity to dark-sector densities and particle masses}
\label{constraints}

We show the forecasted constraints on the axion energy density from CMB-S4 including lensing in Figure~\ref{fig:axions}. We compare $1\sigma$ errors for {\sl Planck} and {\sl Planck}+S4 (where {\sl Planck} is used on a reduced part of the sky as described in Section~\ref{data}) around a fiducial axion fraction $\Omega_a/\Omega_d=2\times 10^{-2}$, and demonstrate the effect of fixing or marginalizing over neutrino mass. We also show forecasted 1 and 2$\sigma$ exclusion lines on Figure~\ref{fig:axions}. In all other error ellipse plots we show $2\sigma$ contours, unless otherwise specified.

\begin{figure}[htbp!]
\begin{center}
\includegraphics[width=0.5\textwidth]{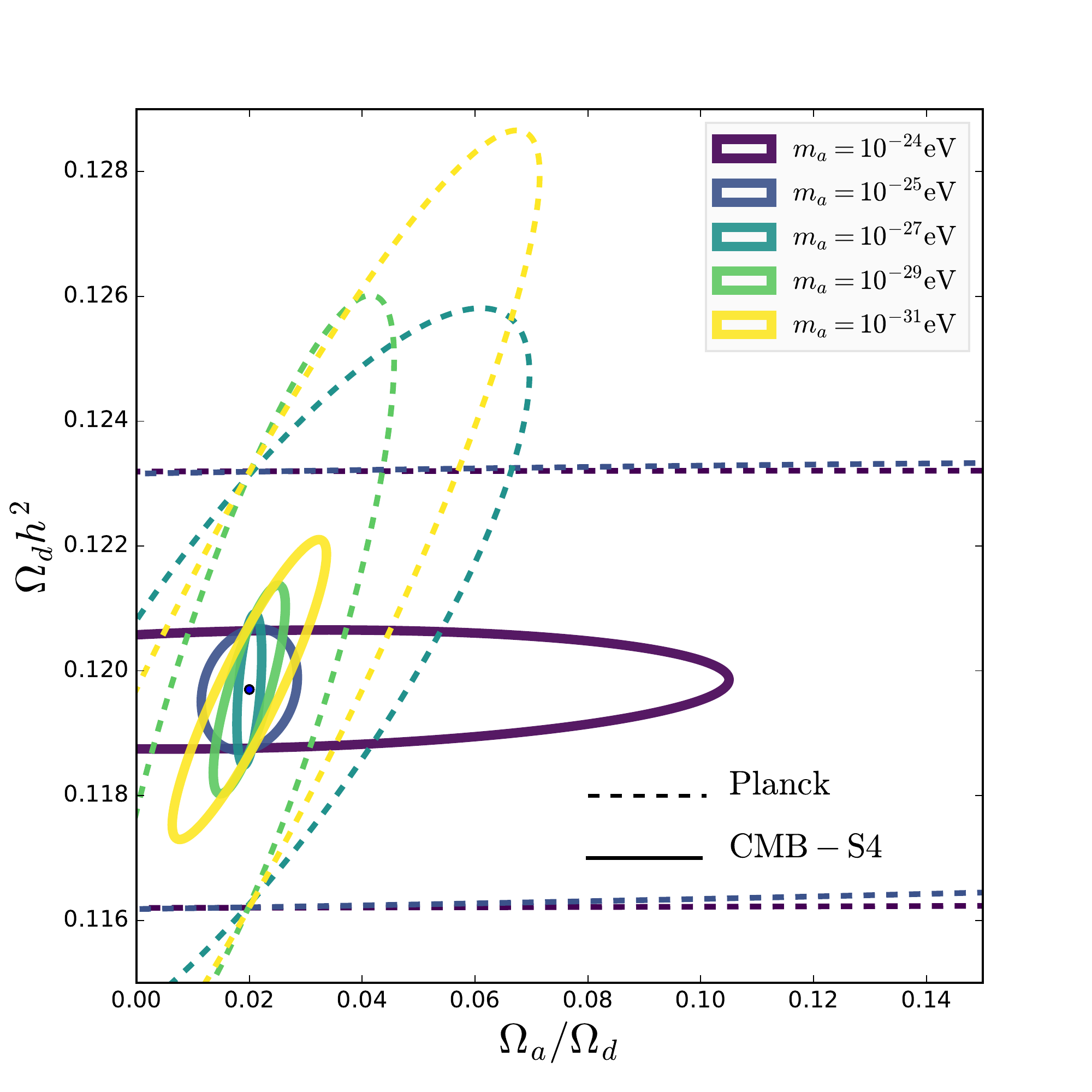}
 \caption{\textbf{The degeneracy between ULAs and CDM for fixed ULA masses:} The fiducial value of the axion fraction, $\Omega_a/\Omega_d=0.02$, is chosen to be consistent with current upper bounds from {\sl Planck}. The dashed lines show forecast constraints based on the {\sl Planck} `Blue Book' \cite{BlueBook2005} sensitivities, and reproduce the constraints using the actual data (see Ref.~\cite{Hlozek:2014lca} for details). The solid lines show constraints for a `CMB-S4-like' survey. At the highest masses considered, $m_a \geq 10^{-24}~\mathrm{eV}$ the axion is completely degenerate with the CDM density: the total dark matter density is well constrained, but the error on the axion fraction becomes larger. The degeneracy direction between axions and CDM rotates as the axion mass changes, with CMB-S4 breaking some strong degeneracies present in {\sl Planck}. In all cases $M_\nu$ has been fixed at its fiducial value, although the constraints in Fig.~\ref{fig:axions} shows that the error on the axion fraction is only degraded for the most degenerate masses in the `fuzzy DM' regime. CMB-S4 would detect a fraction of $\Omega_a/\Omega_d=0.02$ at $>2\sigma$ in the mass range $10^{-30}\,\mathrm{eV} < m_a < 10^{-24.5}\,\mathrm{eV}$.  \label{fig:axion_cdm_degen}}
\end{center}
 \end{figure}

Figure~\ref{fig:axion_cdm_degen} shows the power of a `CMB-S4-like' survey to distinguish ULAs from CDM, by comparing constraints for {\sl Planck}+S4 (solid lines) to constraints assuming only {\sl Planck} specifications (dashed lines). CMB-S4 will not only tighten the constraints on the total DM content, but closes in on the axion parameter space as well. In particular for some masses (most notably $m_a = 10^{-25}\, \mathrm{eV}$), CMB-S4 breaks the degeneracy between ULAs and CDM even at very low axion fraction. CMB-S4 will allow for a multi-$\sigma$ detection of percent level departures from CDM for all masses in the range $10^{-30}\,\mathrm{eV}<m_a<10^{-24}\,\mathrm{eV}$. Thus CMB-S4 presents an ability to test the composition of DM, and thus the CDM paradigm, at the percent level. 

For these most DM-like ULAs $(m_a\geq 10^{-25}~\mathrm{eV})$ the current data (i.e. {\sl Planck}, see Ref.~\cite{Hlozek:2014lca}) do not bound the axion fraction at the percent level. As shown in Figure~\ref{fig:axion_cdm_degen}, {\sl Planck} has essentially no constraining power for $m_a=10^{-24}\,\mathrm{eV}$, when $\Omega_a$ and $\Omega_c$ are totally degenerate. As the axion mass changes, the degeneracy goes from complete (horizontal in this representation), with the error on the total dark content unchanged irrespective of the axion fraction, to one where the axion fraction is tightly constrained (e.g. $m_a=10^{-29}\,\mathrm{eV}$). The degeneracy direction continues to change for the lighter axions as they become more DE-like. 

\begin{figure*}[htbp!]
\begin{center}
$\begin{array}{ll}
\includegraphics[width=1\columnwidth]{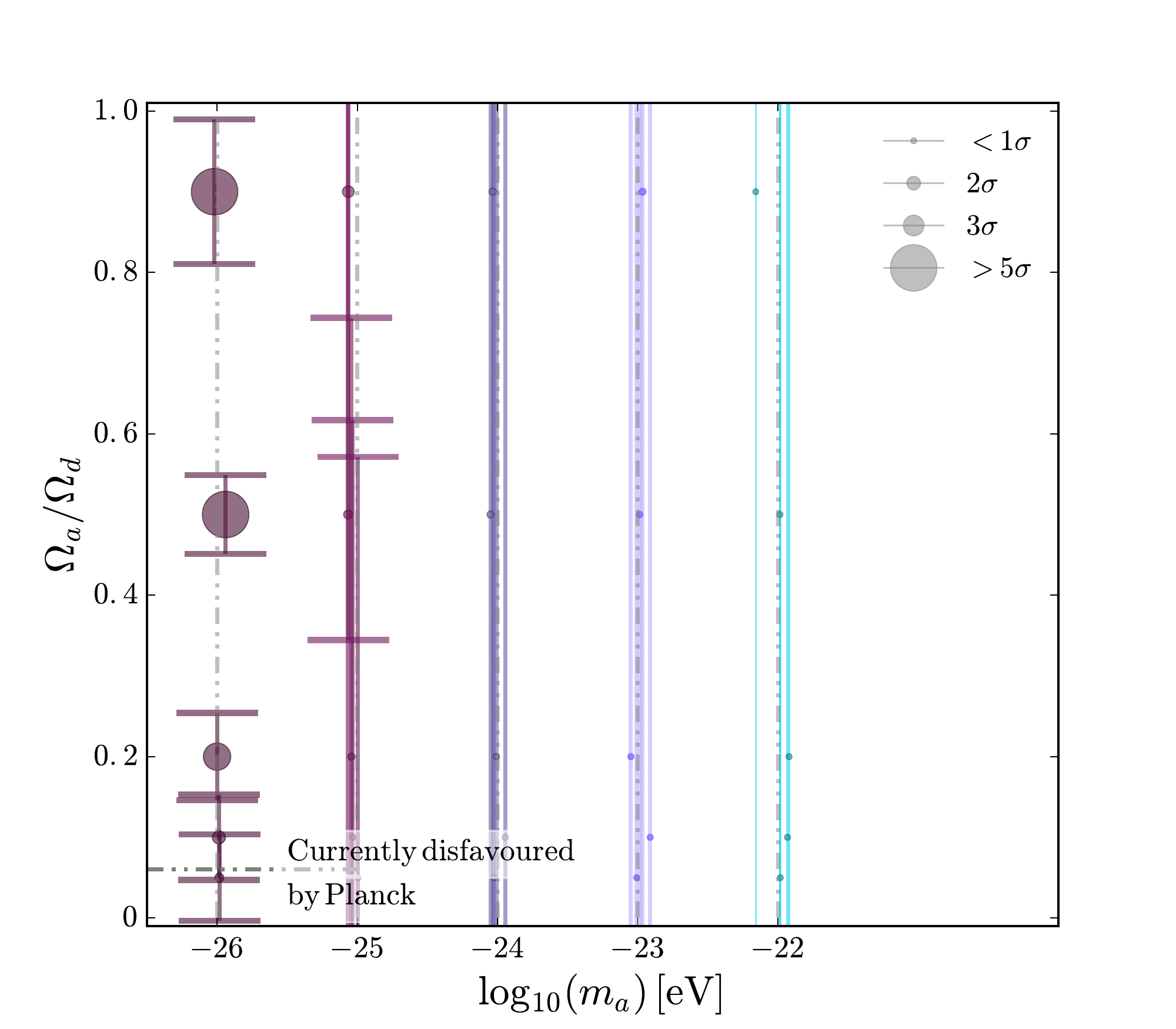} &
\includegraphics[width=1\columnwidth]{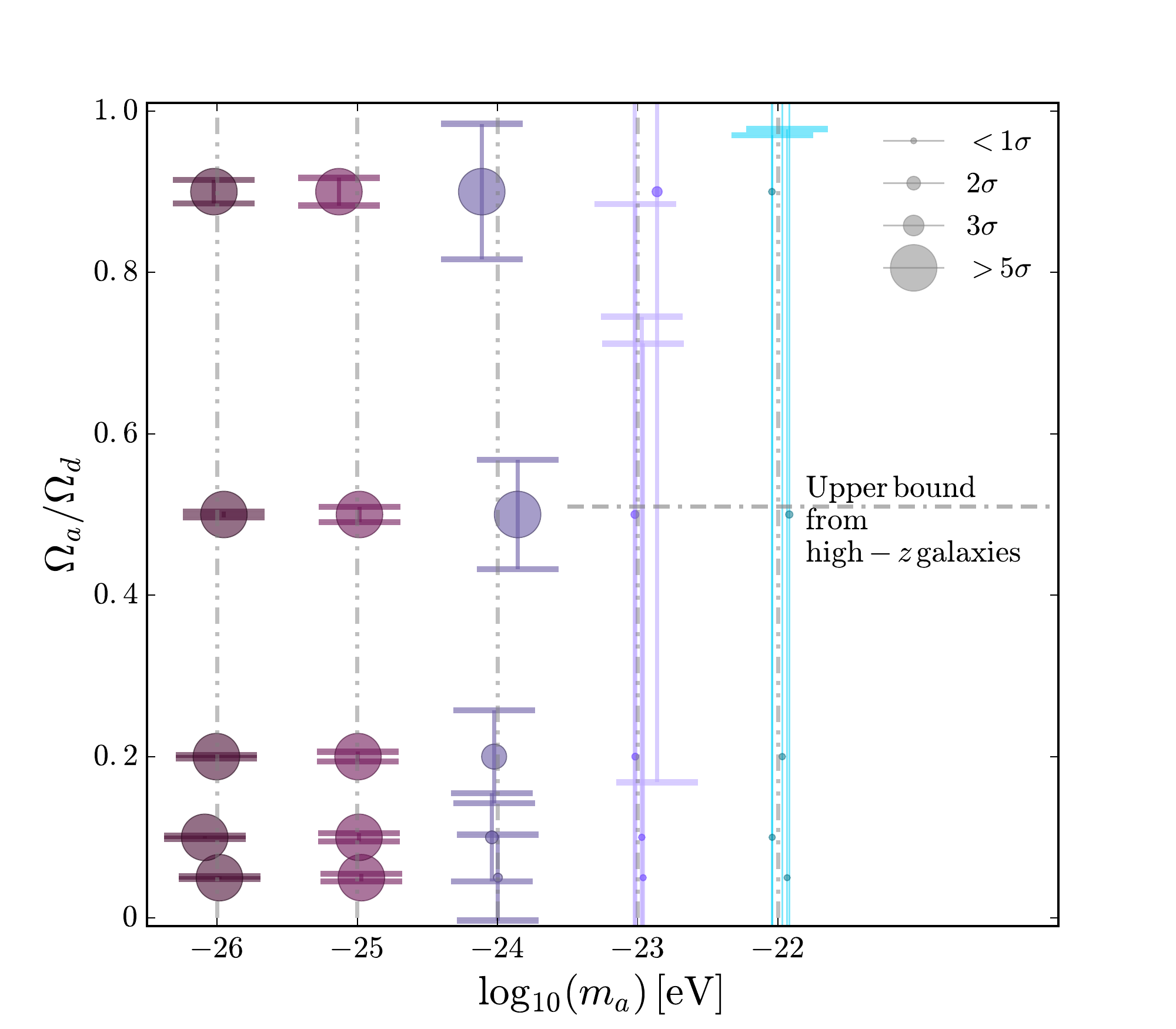}
\end{array}$
 \caption{\textbf{Forecast detection significance of `dark-matter-like' ULAs:} The \textit{left panel} shows the current constraints from a forecast {\sl Planck} survey (which is consistent with the results from Ref.~\cite{Hlozek:2014lca}). The right panel shows the forecast constraints from a `CMB-S4-like' survey over the same fixed masses ranging from $\log_{10}(m_a) = -26~\mathrm{eV}$ to $\log_{10}(m_a) = -22~\mathrm{eV}$. (For ease of viewing a random scatter has been placed in the $x$-direction for each mass, the dashed line gives the central mass value.) The $y-$axis shows the assumed fiducial axion fractions of $0.05, 0.1, 0.2, 0.5$ and $0.9$, with the forecast error on the fraction. The size of the marker is proportional to the significance with which we would detect such a fiducial axion fraction (the size is fixed for all detections $>5\sigma$). For the Planck survey, the constraints are eroded for masses heavier than around $\log_{10}(m_a) = -26~\mathrm{eV}$. CMB-S4 will push this boundary of ignorance by two orders of magnitude. A `CMB-S4-like' survey will allow a detection of an axion fraction as low as $5\%$ at $> 5\sigma$ for $\log_{10}(m_a) = -25~\mathrm{eV}$, and a fraction of $20\%$ at $> 3\sigma$ for $\log_{10}(m_a) = -24~\mathrm{eV}$. \label{fig:frac}}
\end{center}
 \end{figure*}
 
 At the largest axion masses, the near-perfect degeneracy between axions and CDM leaves us without a meaningful upper limit to saturate when choosing fiducial values for $\Omega_{a}/\Omega_{d}$. To test how a `CMB-S4-like' survey might place a tighter upper limit on the fraction of DM made up of ULAs, we instead forecast the significance of a CMB detection of ULAs while varying the fiducial fraction, and consider the detection significance. The results are shown in Figure~\ref{fig:frac}.

We fix the total DM energy density to the fiducial value of $\Omega_dh^2 = 0.1197$ (marginalizing over this and all other parameters) and vary the axion fraction as parameter of interest. We consider a range of fixed axion masses logarithmically spaced between $m_a=10^{-26}\,\mathrm{eV}$ and $m_a=10^{-22}\,\mathrm{eV}$. At each mass we use a range of fiducial fractions $(\Omega_a/\Omega_d = 0.05, 0.1, 0.2, 0.5, 0.9)$ and show the marginalized error on the fraction centred at the fiducial value. In Figure~\ref{fig:frac}, the size of the detection significance (in units of $\sigma$) is illustrated by the size of the marker, and we compare {\sl Planck} to {\sl Planck}+CMB-S4. 

For axion masses of $\log_{10}(m_a)=-24~\mathrm{eV}$ using CMB-S4 an axion fraction as low as $20\%$ could be detected at $>3\sigma$, a vast improvement over {\sl Planck}, which has essentially no constraining power at this mass. We see that {\sl Planck} alone places only $\sim 1\sigma$ limits at high fraction for $m_a=10^{-25}\text{ eV}$ (consistent with the analysis of real data in Ref.~\cite{Hlozek:2014lca}), while this `wall of ignorance' is moved to $m_a=10^{-23}\text{ eV}$ with {\sl Planck}+CMB-S4.

The solid and dashed lines in Figure~\ref{fig:axions} show a different approach to the same issue of setting upper bounds. They show the fiducial models one could \textit{rule out} with $1\sigma$ (dashed) or $2\sigma$ (solid) significance. While the highest mass ULAs, $m\geq 10^{-22}\text{ eV}$, remain completely degenerate with CDM, one could rule out a fraction of $>15\%$ at $2\sigma$ confidence at $m_a=10^{-24}\,\mathrm{eV}$ and one could rule out an axion fraction of $>64\%$ at $1\sigma$ confidence at $m_a=10^{-23}\,\mathrm{eV}$. Figs~\ref{fig:axions} and~\ref{fig:frac} show how CMB-S4 could improve the lower limit on DM particle mass from the CMB alone by approximately 2 orders of magnitude compared with {\sl Planck}. 

\begin{figure}[htbp!] 
\begin{center} 
\includegraphics[width=0.5\textwidth, height=0.47\textwidth]{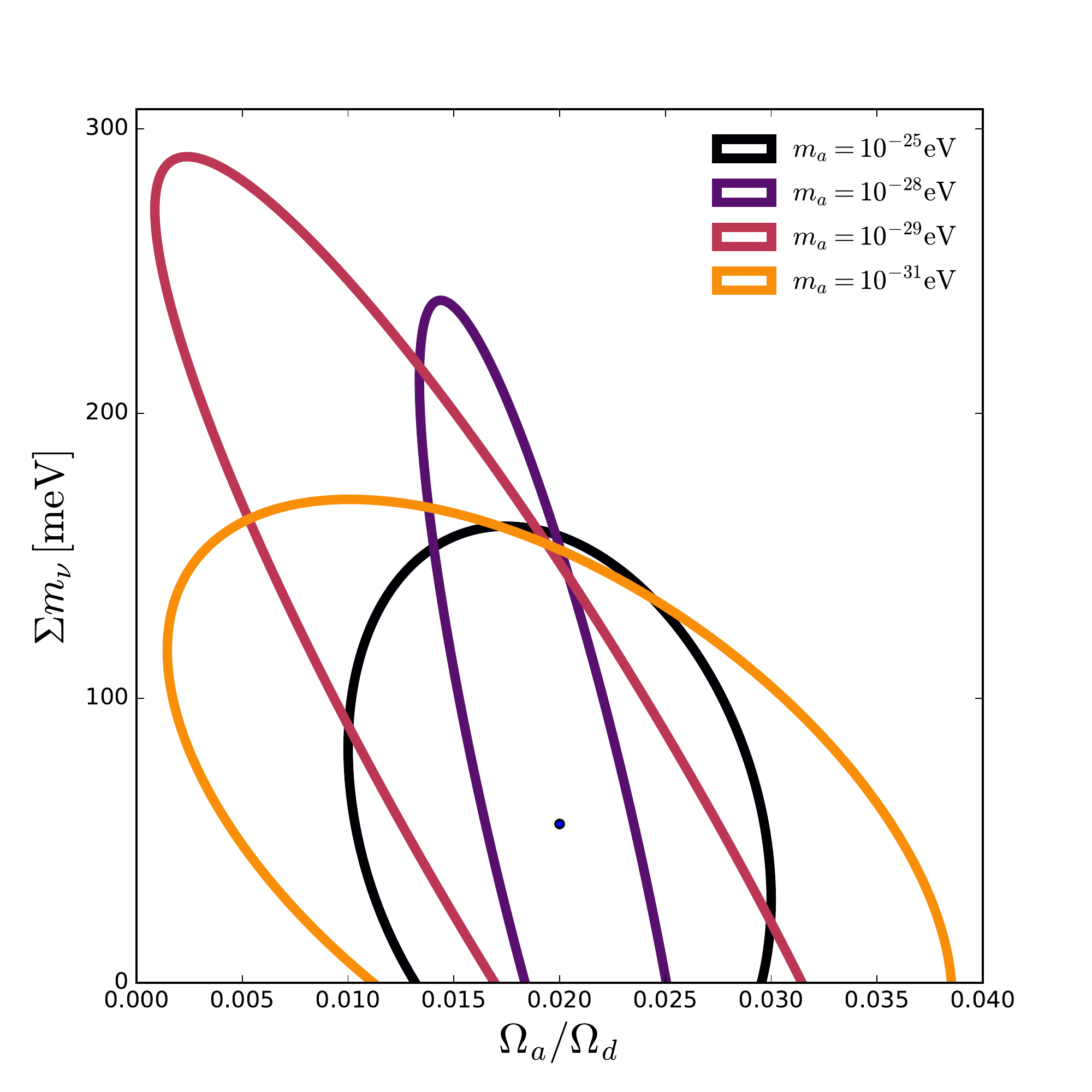} 
 \caption{\textbf{The degeneracy of ULAs with massive neutrinos:} The lighter ULAs show a significant degeneracy with neutrino mass for a `CMB-S4-like' survey, as summarized in Table~\ref{table:net}. The error bars increasing towards lighter mass - as these DE-like ULAs are less constrained with future data. Adding a prior on the expansion rate will reduce the errors on these parameters, as shown in Figure~\ref{fig:axions_prior} \label{fig:axions_mnu}}.
 \end{center}       
 \end{figure}
\begin{figure}[htbp!] 
\begin{center}
\includegraphics[width=0.45\textwidth]{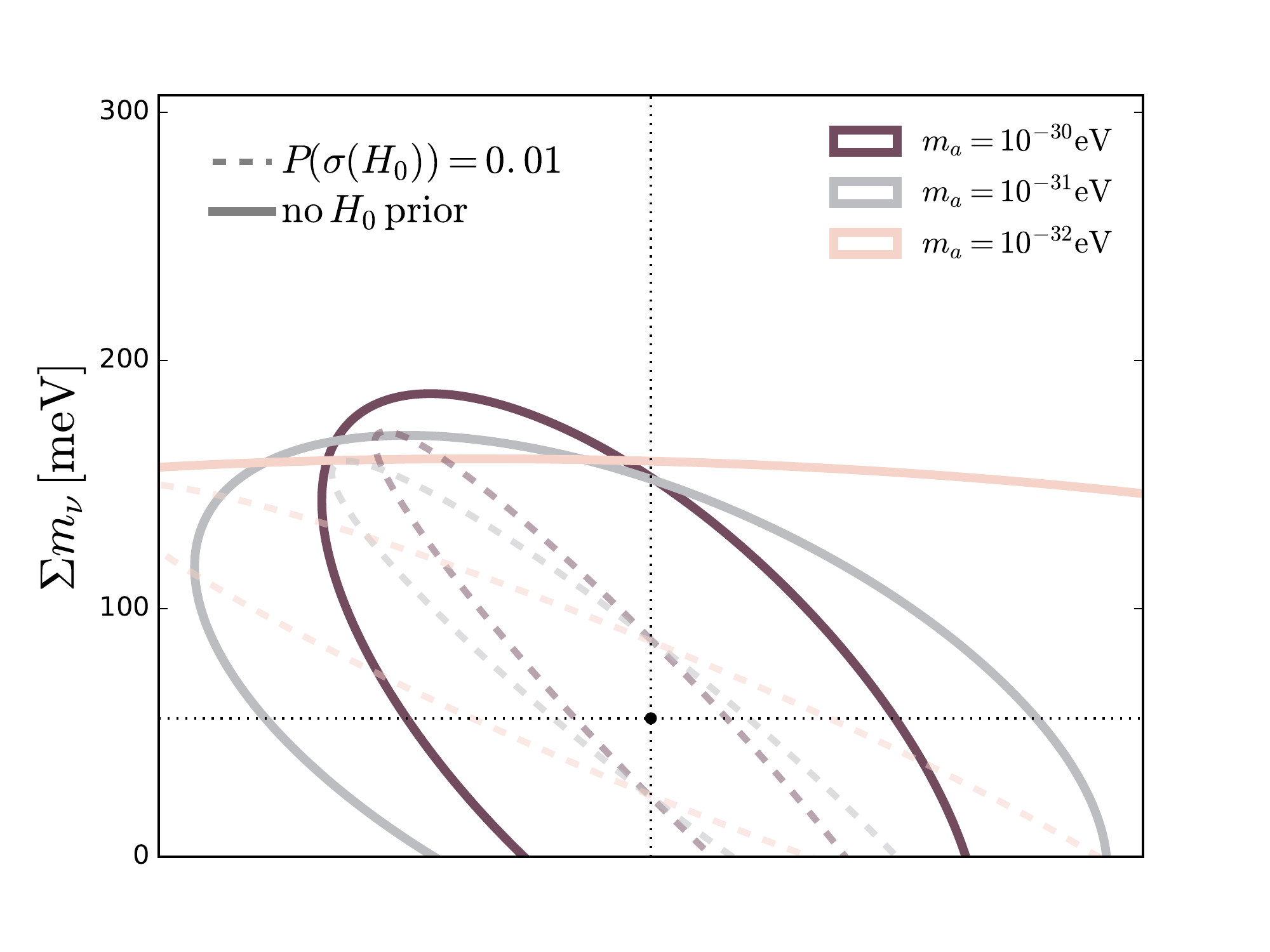} \\
\includegraphics[width=0.45\textwidth]{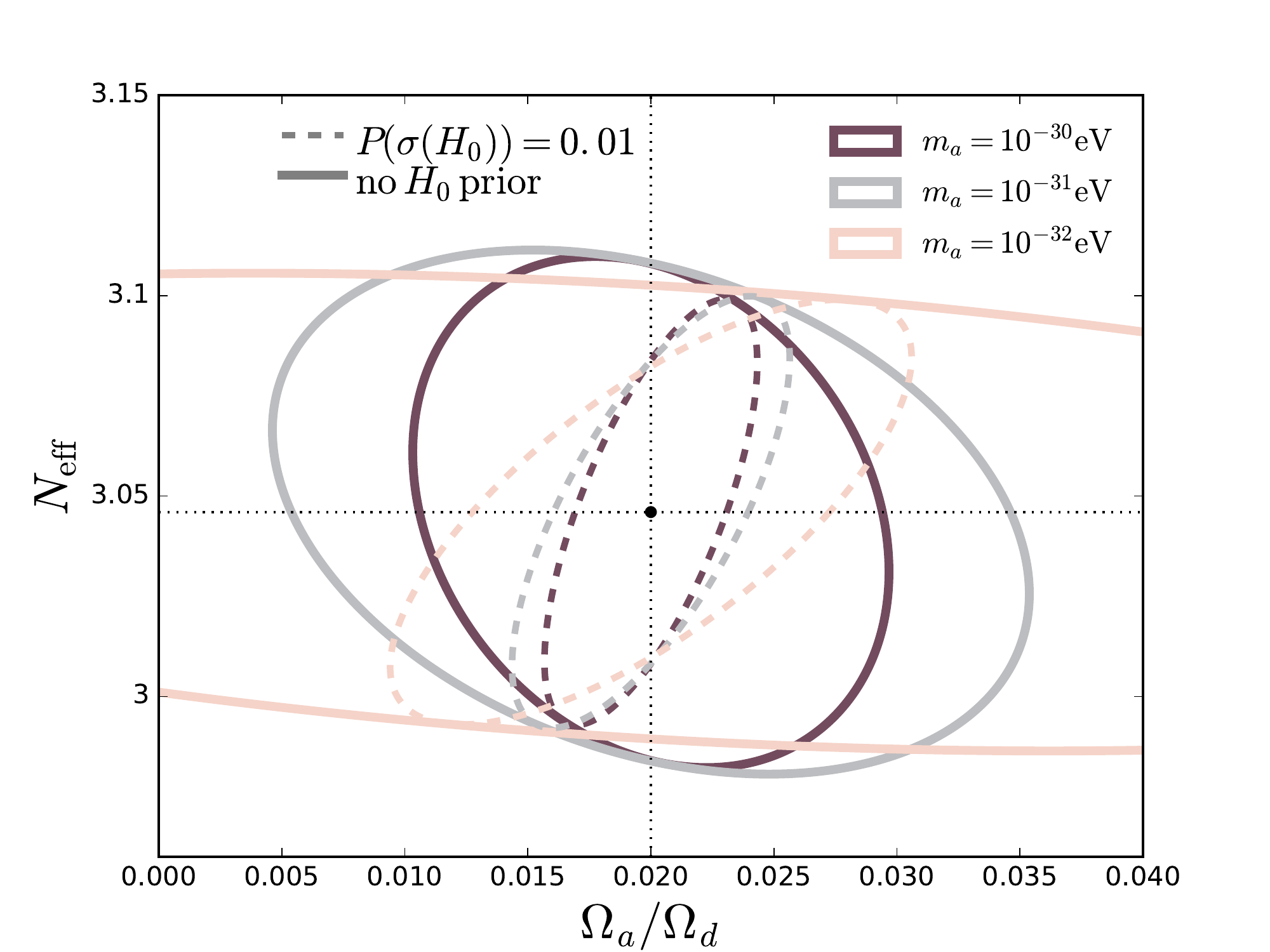} 
 \caption{\textbf{Priors on the expansion rate improve constraints on the lightest ULAs:} The degeneracies between the ULAs with mass $m_a < 10^{-30}~\mathrm{eV}$ and massive neutrinos (\textit{top panel}) and massless species (\textit{bottom panel}) are shown for a `CMB-S4-like' experiment (as specified in Table~\ref{table:net}), with the solid lines showing the constraints without any additional prior on the Hubble constant (there is some repetition with the left panel here and in Figure~\ref{fig:axions_mnu}). The dashed lines show the improvement when adding a prior of 1\% on $H_0$ from a `DESI-like' experiment \cite{Font-Ribera:2013rwa}. \label{fig:axions_prior}}
 \end{center}       
 \end{figure}

The degeneracies of the ULAs with other cosmological parameters, such as $N_\mathrm{eff}$ or $\Sigma m_\nu$, also varies depending on the axion mass (see Figs.~\ref{fig:axions_mnu} and~\ref{fig:axions_prior}). As described already, DE-like ULAs with masses around $10^{-33}~\mathrm{eV}$ change the late-time expansion rate and therefore the sound horizon, changing the location of the acoustic peaks. This has degeneracies with the matter and curvature content. Heavier ULAs ($m_a \gtrsim 10^{-26}~\mathrm{eV}$) affect the expansion rate in the radiation era and reduce the angular scale of the diffusion distance, leading to a boost in the higher acoustic peaks, which has a degeneracy with $N_\mathrm{eff}$. 

Consider the degeneracy between $\Sigma m_\nu$ and axion fraction, varying the axion mass (Fig.~\ref{fig:axions_mnu}). Certain axion masses are more degenerate with the fiducial neutrino model than others, making for example, a $m_a = 10^{-29}\, \mathrm{eV}$ axion more prone to masquerading as a massive neutrino than an axion of mass $m_a = 10^{-25}\, \mathrm{eV}$ (for a $m_a=10^{-29}\,\mathrm{eV}$ axion, the error on $\Sigma m_\nu$ is halved relative to the $m_a = 10^{-25}\, \mathrm{eV}$ case). The degeneracy is not total, however, and we will still be able to make a significant detection of a small axion fractions, using CMB-S4. Additionally, this degeneracy can be broken by local measurements of $H_{0}$.
 
 As a test of how $H_0$ measurements can change constraints on the lightest ULAs, we added a prior of $1\%$ on $H_0$ to our forecasts. Current local measurements provide a $2-3\%$ constraint \cite{Riess:2016jrr}, while future efforts like DESI \cite{Font-Ribera:2013rwa} will provide roughly percent-level measurements from BAO. The addition of this prior changes the error on the axion fraction for an axion of mass $m_a = 10^{-32}\, \mathrm{eV}$  (assuming a fiducial fraction of 0.02) from $0.03$ to $0.005$ - allowing a $>4\sigma$ detection of the axion fraction even at the lowest masses. Local measurements of $H_0$ constrain the effects that these ULAs have on the low-$z$ expansion rate.

Figure~\ref{fig:axions_prior} shows how adding a $H_0$ prior to the precise measurement of the temperature and polarization power with CMB-S4 leads to an improvement in the error on $\Omega_a/\Omega_d$ at low ULA mass ($m_a\leq 10^{-30}\text{ eV}$). We show how the $H_0$ prior affects ULA degeneracy with $\Sigma m_\nu$ (left panel) and $N_\mathrm{eff}$ (right panel). In both cases the inclusion of a $H_0$ prior does not have a large effect on the error in the neutrino parameters ($\Sigma m_\nu$ or $N_\mathrm{eff}$), but it greatly reduces the degeneracy between light ULAs and neutrinos. The $H_0$ prior reduces the uncertrainty on $\Omega_a/\Omega_d$ by a factor of $\approx 3$ where both $\Sigma m_\nu$ and the axion fraction are varied, and a factor of $\approx 5$ when $N_\mathrm{eff}$ is varied with the axion fraction.

\begin{figure}[htbp!] 
\begin{center} 
\includegraphics[width=0.5\textwidth]{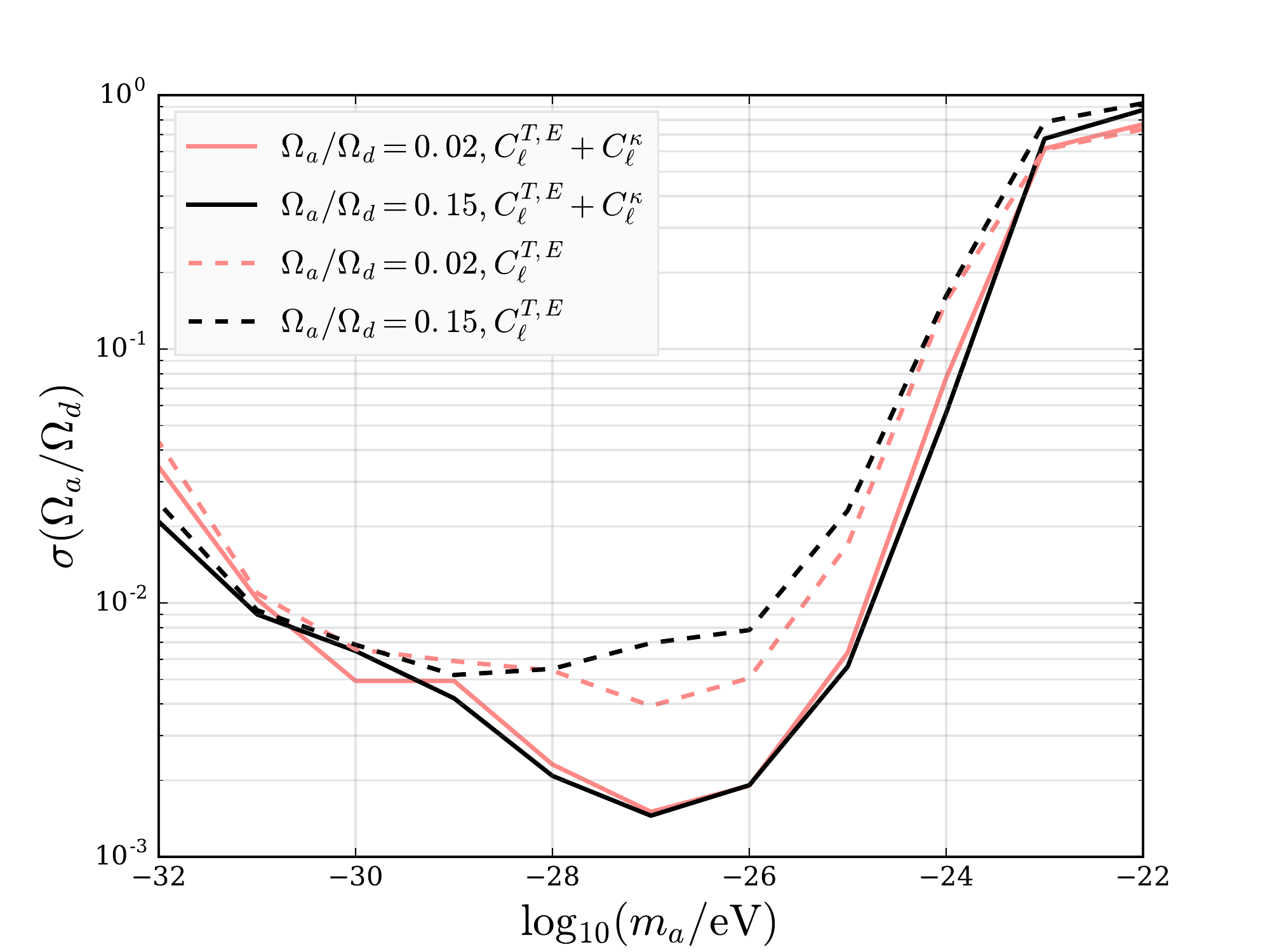}\\ [0.0cm]
 \caption{\textbf{Constraints on the axion fraction with and without lensing:} For a `CMB-S4-like' survey, the $1\sigma$ marginalized error bar on the axion fraction, $\Omega_a/\Omega_d$, for the ranges of masses considered: $10^{-32}<m_a< 10^{-22}\text{ eV}$. For masses $\log(m_a/\mathrm{eV}) > -28$, lensing more than halves the error bar for the same survey parameters where the lensing deflection is not included. The improvement is also sensitive to the fiducial model of ULAs assumed. This is particularly relevant given that for the heaviest masses the ULAs are currently indistinguishable from a standard DM component. \label{fig:lensing_error}}
 \end{center}       
 \end{figure}

The power of CMB-S4 lensing to break the degeneracy between ULAs and CDM is shown in Figure~\ref{fig:lensing_error}, which compares the error bar with and without adding in the lensing deflection measurements (solid to dashed line comparison) for different fiducial models. The largest reduction in the error including lensing deflection measurements comes in the mass range $10^{-29}\,\mathrm{eV} <m_a < 10^{-24}\,\mathrm{eV}$. 

For CMB-S4 and an axion mass of $m_a = 10^{−26}\,\mathrm{eV}$, the percent-level measurement of the lensing power at multipoles $\ell > 1000$ leads to an improvement in the uncertainty on the axion energy density of a factor of eight relative to case where lensing information is excluded. Lensing also plays a key role in the ability of CMB-S4 to improve constraints on ULAs in the range $10^{-24}\,\mathrm{eV} <m_a < 10^{-22}\,\mathrm{eV}$. 
 
\subsection{Survey optimization}\label{survey_optimize}

The specifications of a `CMB-S4-like' survey are shown in Table~\ref{table:net}. One might ask what survey parameters might be most suitable to maximise constraints on the axion parameter space.

\begin{figure}[htbp!]
\begin{center}
\includegraphics[width=0.5\textwidth]{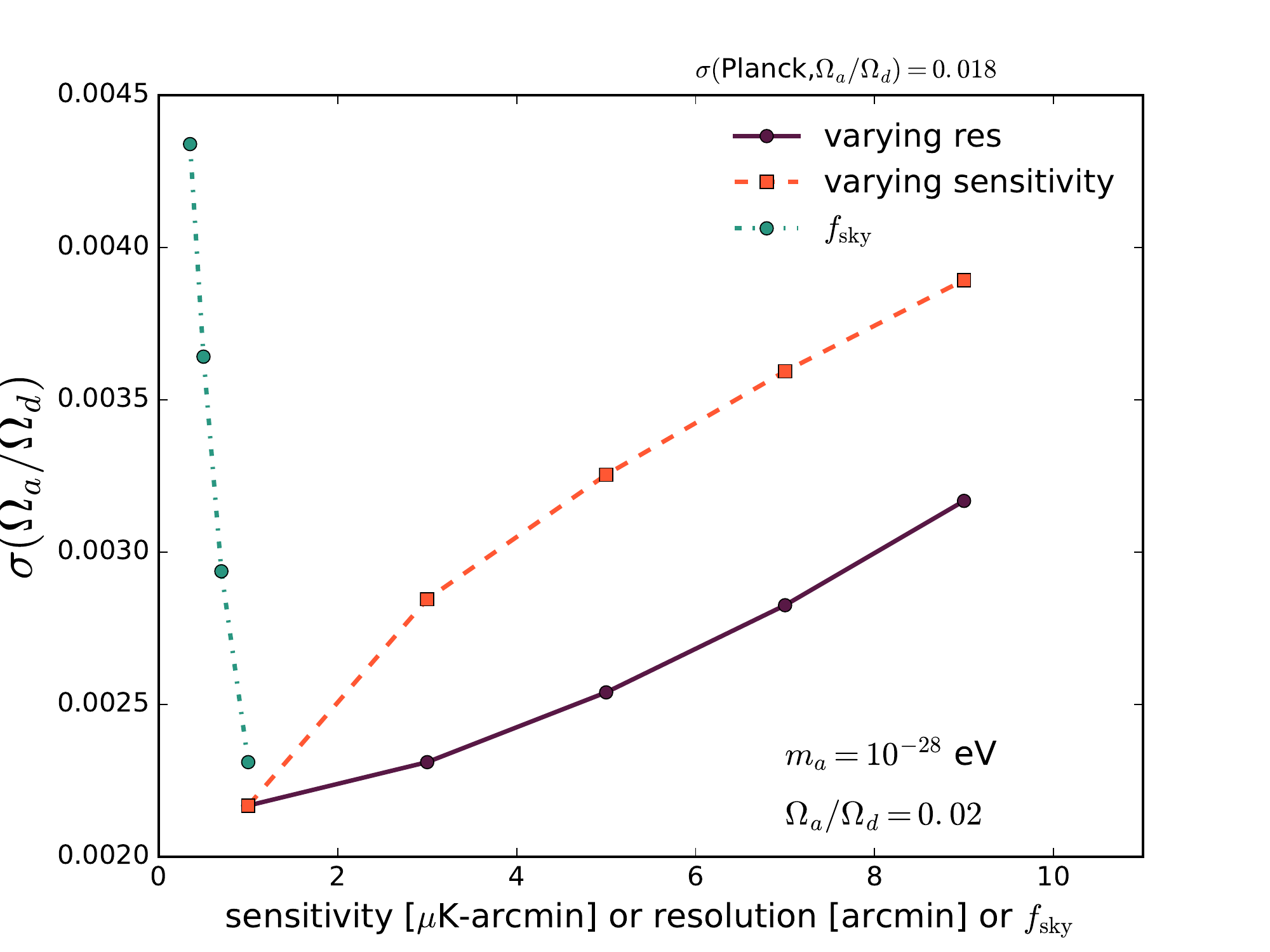}
 \caption{\textbf{Constraints on the axion fraction as a function of survey parameters:} We vary the resolution and sensitivity for a range of `CMB-S4' survey parameters, around the baseline parameters of $1~\mu$K-arcmin, a resolution of 1 arcmin and a baseline sky fraction for CMB-S4 of $f_\mathrm{sky}=0.4,$ which is supplemented with a correspondingly reduced area of the {\sl Planck} sky. The error degrades slowly with worse resolution (solid line) and sensitivity (dashed line). The dot-dashed line shows the constraints for fixed observing time, changing the fraction of sky and accordingly modifying the sensitivity of the `CMB-S4-like' survey (and the amount of sky covered in corresponding {\sl Planck} maps). Since the ULAs affect the small-scale damping tail and the lensing deflection most strongly, moving to small, sensitive patches of the sky increases the error on the axion density (as opposed to having a fixed value of $f_\mathrm{sky}$ but pushing for lower instrumental noise levels). Conversely, tripling the beam size does not have a strong effect on the error on the axion fraction. \label{fig:s4optimise}}
\end{center}
 \end{figure}

We show the results of some choices for the beam size and noise sensitivity in Figure~\ref{fig:s4optimise}. In each case we either vary the beam and sensitivity separately (solid and dashed lines), or we change the sky area at fixed 1 arcminute beam resolution, while adjusting the sensitivity assuming fixed total number of detectors and observing time. In the case where we \emph{reduce} the amount of sky observed by S4, we adjust the correponding area used from the {\sl Planck} satellite to include the fraction \emph{not} observed by S4. This is shown in the Figure with a dot-dashed line. 

As discussed in Section~\ref{observables}, ULAs affect largely the high-$\ell$ damping tail of the CMB lensing deflection power, and so improvements in the noise properties at small angular scales tightens constraints on ULAs. 
Moving to small, deep patches of the sky does not reduce the error: to constrain ULAs we need larger sky area given a total noise budget.

\section{Conclusions \label{conclusion}}
We live in the age of precision cosmology. Future experiments like the proposed CMB-S4 will significantly improve constraints on the composition of the dark sector. We have shown in detail how this is achieved in the case of ultra-light axions, including degeneracies with dark radiation and massive neutrinos. CMB-S4 will move the wall of ignorance for the heaviest axion candidates from $m_a=10^{-26}~\mathrm{eV}$ to $m_a=10^{-24}~\mathrm{eV}$ (detection with an axion fraction of 20\% at $>3\sigma$). 

The lower limit on the dominant DM particle mass will be increased from $m_a=10^{-25}~\mathrm{eV}$ to $m_a=10^{-23}~\mathrm{eV}$ ($1\sigma$ constraints rule out large fractions). This begins to make contact with the much more systematic-laden upper bounds on the axion mass and fraction from high-$z$ galaxies and reionization: $\Omega_a/\Omega_d<0.5$ for $m_a=10^{-23}\text{ eV}$ and $m_a\gtrsim 10^{-22}\text{ eV}$ for the dominant component~\cite{Bozek:2014uqa,Schive:2015kza,Sarkar:2015dib}. This value approaches the mass range needed to explain dwarf galaxy cores and missing Milky Way satellites~(e.g. Refs.~\cite{Hu:2000ke,Marsh:2013ywa,Schive:2014dra,Marsh:2015wka}).

Perhaps more impressively, the constraints on the axion energy density at intermediate mass could improve by an order of magnitude. CMB-S4 could detect an axion fraction as low as $0.02$ at $>13\sigma$ for an axion mass of $10^{-27}\, \mathrm{eV}$.

Given the power of these future efforts, it will be possible to probe the degeneracies between ULAs and other potential DM components, such as massive neutrinos, and light species such as massless sterile neutrinos.

Improved independent constraints on measurements of the expansion rate (through measurements of the Hubble constant, for example) will improve sensitivity to the lightest, DE-like axions, and reduce the degeneracy between these species and both $\Sigma m_\nu$ and $N_\mathrm{eff}$. Even when marginalizing over the neutrino mass, the error on the axion fraction for a mass of $m_a = 10^{-32}\, \mathrm{eV}$ improves by a factor of three with a prior on the expansion rate.

As $\Omega_a\propto f_a^2$ the improved sensitivity to the axion energy density improve the axion decay constant which could be detected from $10^{17}\mathrm{ GeV}$ with {\sl Planck} to $10^{16}\mathrm{ GeV}$ with CMB-S4 (over the relevant range of ULA masses). The improved sensitivity to $f_a$ will begin to test the predictions of the string axiverse scenario~\cite{Arvanitaki:2009fg}. 

Axions are a well motivated dark matter candidate, and future CMB experiments suggest an exciting opportunity to explore the rich complexity of their parameter space, moving towards sub-percent level sensitivity to the axion energy density or a $10\sigma$ detection if current limits to $\Omega_{a}$ are saturated by the true axion density, all over for a wide range of masses. As a spectator field during the inflationary era, axions would also carry isocurvature perburbations (see Ref. \cite{Hlozek:2014lca} and references therein), leading to distinct imprints on CMB observables and providing a unique new lever arm on the inflationary energy scale, which is otherwise only accessible through measurements of primordial CMB B-mode polarization \cite{Marsh:2014qoa}. In future work, we will extend {\sl Planck} constraints and CMB-S4 forecasts to include the impact of isocurvature.

Unraveling the mystery of dark matter is an important goal for cosmology in the coming decades. The axion represents the lowest mass DM-candidate, and a `CMB-S4-like' survey will help identify (or rule out) these models of DM. Constraints on the light, DE-like axions are improved by independent measurements of the expansion rate of the Universe, thereby probing our knowledge of the cosmological constant, quintessence, and cosmic acceleration in general.

In this work, we have illustrated that future CMB experiments will shed new light on the nature or existence of the axion and usher axiverse cosmology into a new era.

\acknowledgements{We thank Pedro Ferreira, Wayne Hu, Alexander Mead and Simeon Bird for useful discussions. RH acknowledges funding from the Dunlap Institute. DJEM acknowledges funding from the Royal Astronomical Society. DG is funded at the University of Chicago by a National Science Foundation Astronomy and Astrophysics Postdoctoral Fellowship under Award  AST-1302856. This work was supported in part by the Kavli Institute for Cosmological Physics at the University of Chicago through grant NSF PHY-1125897 and
an endowment from the Kavli Foundation and its founder Fred Kavli. RA and JD are supported by ERC grant 259505, and EC is funded by the STFC Ernest Rutherford Fellowship.}
\appendix
\section{\textsc{AxionCAMB} code}\label{code_appendix}
In this work, we use a specially modified version of the CMB Boltzmann code \textsc{camb}, called \textsc{AxionCAMB}, in order to compute the primary CMB anisotropy power spectra $C_{l}^{\rm TT}$, $C_{l}^{\rm EE}$, and $C_{l}^{\rm TE}$, as well as the lensing-convergence power spectrum $C_{l}^{\kappa \kappa}$.\footnote{A related modification to \textsc{class}~\cite{2011arXiv1104.2932L} by other authors was reported recently in Ref.~\cite{2015arXiv151108195U}, but this has not been publicly released. The results of Ref.~\cite{2015arXiv151108195U} seem qualitatively similar to \textsc{axionCAMB}, though a formal code comparison would be useful.} Weak lensing of the CMB smears out the peaks in the primary CMB power spectra and introduces non-Gaussian features into the CMB temperature and polarization fields \cite{Hu:2001kj}; the resulting effect on temperature/polarization $4$-pt functions can be used to reconstruct $C_{l}^{\kappa \kappa}$, as discussed in Refs.~\cite{Goldberg:1999xm,Zaldarriaga:1998te,Hu:2001kj,Okamoto:2003zw,Lewis:2006fu}. \textsc{AxionCAMB} is used to compute the matter power spectrum $P_{m}(k)$ and then $C_{l}^{\kappa \kappa}$ , which is necessary for our forecast. \textsc{AxionCAMB} has already been used to obtain the results of Ref.~\cite{Marsh:2014qoa,Hlozek:2014lca}, but since that work, we have improved the code, and present it here for public use. The code may be downloaded on the \textsc{GitHub} repository.\footnote{\url{http://github.com/dgrin1/axionCAMB}.} We welcome comments and useful additions/improvements to the code.

After cosmological parameters (including the ULA parameters $m_{a}$ and $\Omega_{a}$) are specified, \textsc{AxionCAMB} begins by solving the coupled Friedmann/Klein-Gordon system for the homogeneous ULA field $\phi_{0}$, where the Klein-Gordon equation in an expanding universe is given by
\begin{equation}
\ddot{\phi}_{0}+2\mathcal{H}\dot{\phi}_{0}+m_{a}^{2}a^{2}\phi_{0}=0.\end{equation} We use a higher-order Runge-Kutta method as described in Ref. \cite{fehlberg}. At early times, this is used to obtain the axion equation of state $w(a)$ and adiabatic sound-speed $c_{\rm ad}(a)$. These quantities can be used to evolve axion energy and pressure perturbations in concert with the usual \textsc{camb} variables at early times, when $a\ll a_{\rm osc}$ ($m_{a} \ll a \mathcal{H}$), using the generalized dark matter formalism of Ref. \cite{Hu:1998kj}: this is not an approximation, but just a useful recasting of the perturbed Klein-Gordon + Einstein equation system at early times. 

At late times ($a\gg a_{\rm osc}$) these equations become stiff, and so we switch to the approximation $w=0$ ($\rho_{a}\propto a^{-3}$, $P_{a}\simeq 0$), with perturbations evolved in the WKB approximation for ULAs, justified in Refs. \cite{Brandenberger:1984jq,Khlopov:1985jw,Peebles:1987ek,Nambu:1989kh,Ratra:1990me,Ratra:1991kz,Hu:2000ke,Hwang:2009js,Arvanitaki:2009fg,Marsh:2010wq,2013CoPhC.184.1339M,2012PhRvD..86h3535P}. 

In the WKB approximation, ULAs may be treated as a fluid with a scale-dependent sound speed:
\begin{equation}
c_{s}^{2}=\frac{\frac{k^{2}}{4m_{a}^{2}a^{2}}}{1+\frac{k^{2}}{4m_{a}^{2}a^{2}}}.
\end{equation}This approximation captures the uncertainty-principle driven suppression of axion perturbation growth without requiring a code that resolves the short oscillation timescale ($\sim m^{-1}$) of ULAs. This treatment allows us to follow perturbations continuously from the slowly rolling to fuzzy dark matter regime for ULAs of any mass, and to explore the parameter space of both dark-energy like and dark-matter like ULAs, as described in Ref. \cite{Hlozek:2014lca}. We have performed a variety of numerical tests (described in Ref. \cite{Hlozek:2014lca}) to confirm that this approximation is sufficiently accurate for analysis/forecasting of realistic CMB and galaxy-clustering data for the foreseeable future. Further details of the implementation are discussed in Ref.~\cite{Hlozek:2014lca}.

Since the work of Ref. \cite{Hlozek:2014lca}, we have improved the scalar-field evolution module of \textsc{AxionCAMB} to properly include the effect of massive neutrinos, using the routines/expressions for time-evolution of the massive neutrino energy-density implemented in \textsc{camb} and discussed in Refs. \cite{cambnotes,Shaw:2009nf}. Note that in \textsc{AxionCAMB} we have also included the radiation energy-density in the closure relation for cosmological densities $1-\Omega_{k}=\Omega_{b}+\Omega_{c}+\Omega_{a}+\Omega^{m}_\nu+\Omega_{\nu}^{r}+\Omega_{\gamma}$, where $\Omega_{\nu}^{m/r}$ is the cosmological energy density in massive/massless neutrinos.

\section{Nonlinear modeling}\label{nonlinear_appendix}

The functional form of the \textsc{halofit} power spectrum is based on the halo model~\cite{Cooray:2002dia}. \textsc{halofit} and the halo model apply only to matter collapsed into halos. In the halo model, this is accounted for using the collapsed mass fraction (from Press-Schechter) and the clustering of the ``smooth component,'' which reduces the halo model power to the linear power if the collapsed fraction is zero~\cite{Smith:2011ev}. 
\begin{figure}[htbp!]
\begin{center}
\includegraphics[width=0.9\columnwidth]{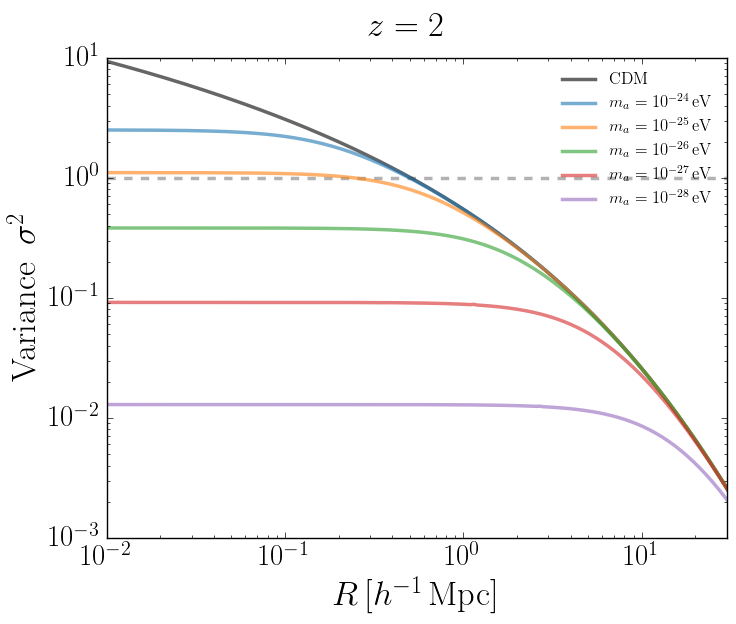}
 \caption{\textbf{Determination of $m_{\rm lin}$ appropriate for our fiducial lensing models:} The axion energy density is $\Omega_a h^2=0.001$. Only masses $m_a\geq m_{\rm lin}$ have density perturbations which are non-linear for $z\geq 2$, with non-linear scale $k_{\rm nl}\approx 1 \, h\text{ Mpc}^{-1}$. The CMB lensing power at high-$\ell$ is dominantly sourced by higher redshifts, and larger scales. For the purposes of lensing forecasts, it is thus consistent to model ULAs with $m_a<m_{\rm lin}$ as linear, following standard treatments of DE and neutrinos. \label{fig:ax_var_z2}}
\end{center}
 \end{figure}
In \textsc{halofit}, this is accounted for by setting the power to linear if the variance, $\sigma^2(R)$, on length scales, $R$, of interest satisfies $\sigma^2(R)<1$ . Since ULAs exhibit suppressed structure formation compared to CDM, the lightest ULAs have collapsed fraction of close to zero even at $z=0$. Furthermore, \textsc{halofit} and the halo model treat all matter components equivalently. We must decide how to include the lightest, sub-dominant, ULAs in the computation of the non-linear ratio, $\mathcal{R}_{\rm nl}(k,z)$. 
\begin{figure}[htbp!] 
\begin{center} 
\includegraphics[width=0.5\textwidth]{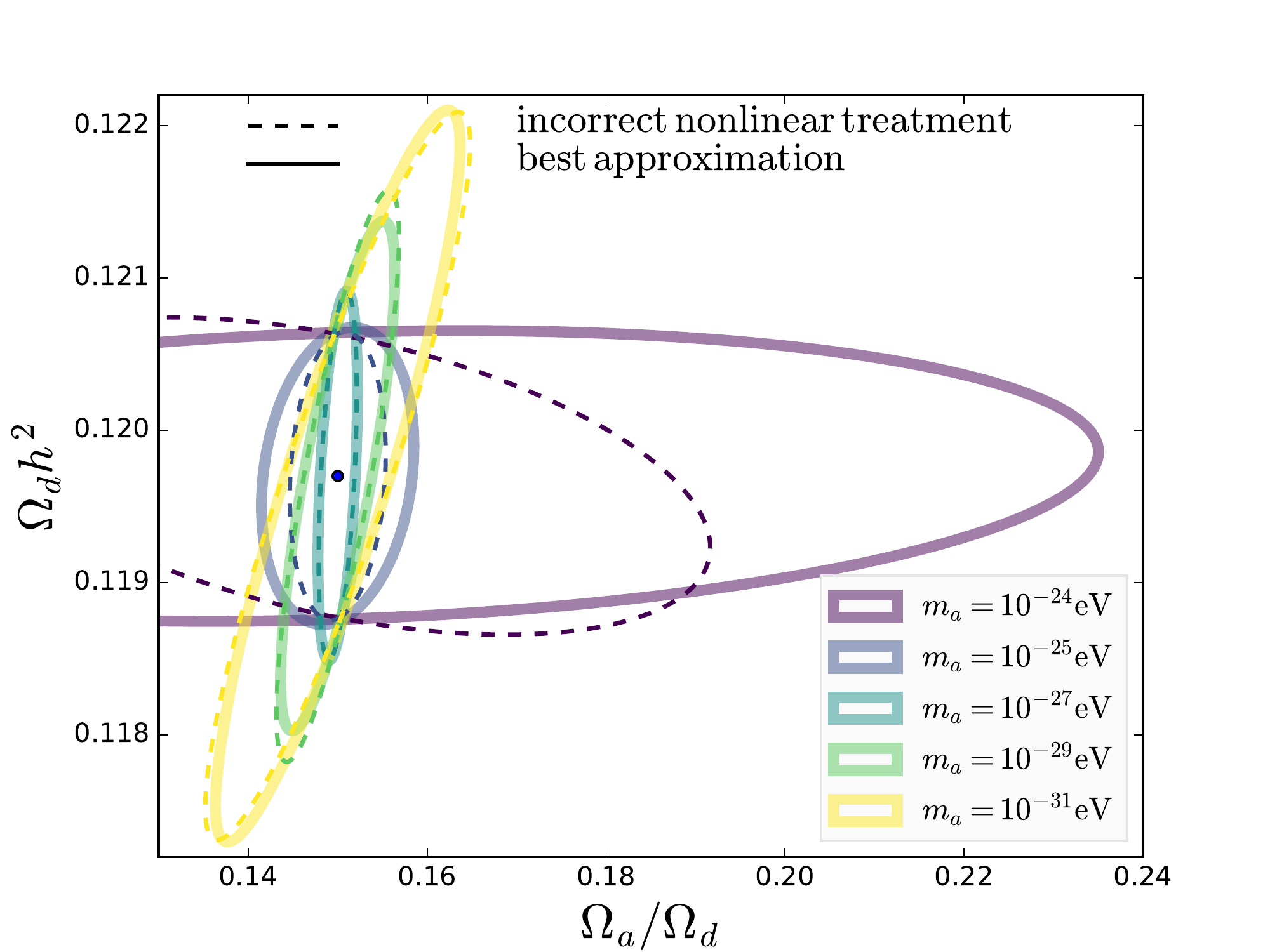} 
 \caption{\textbf{The effect of nonlinear halo modeling on axion constraints:} Marginalized $2\sigma$ contours showing the degeneracy between ULAs and the cold dark matter energy density for a range of masses in a CMB-S4-like survey. The difference between solid and dashed lines is only the nonlinear halo modeling. For the lightest axions, the effect is small, however for $m_a \geq 10^{-25}$ eV it can lead to spuriously tight constraints on the axion fraction. \label{fig:axions_nonlinear}}
 \end{center}       
 \end{figure} 
First, consider the lightest DE-like ULAs. For these ULAs, we adopt a simple criterion for the non-linear modeling, by analogy to \textsc{camb}'s treatment of DE models (and the strict equivalence between ULAs and quintessence as $m_a\rightarrow 0$). We choose to only include ULAs in the ``non-linear matter'' [i.e. in $P(k)$ used to compute the non-linear ratio] if $\sigma_a^2(R\rightarrow 0,z)>1$, where $\sigma_a^2$ is the variance in the axion power spectrum. 

Ideally, this criterion should be computed for every redshift $z<z_{\rm rec}$ and for all combinations of cosmological parameters separately. Instead, for simplicity in the current study, we make a hard cut on axion mass:
\be
m_{\rm lin}= 10^{-25}\text{ eV} \, .
\label{eqn:ma_linear}
\ee
ULAs with $m_a< m_{\rm lin}$ are treated passively in Eq.~(\ref{eqn:psi_non-lin}), i.e. are included in $\mathcal{P}_{\Psi,\rm lin}$ but do not appear in $\sigma_m^2$ used to compute $\mathcal{R}_{\rm nl}$. 

The cut, Eq.~(\ref{eqn:ma_linear}), is appropriate for CMB lensing forecasts with fiducial models allowed by the constraints imposed by {\sl Planck}-2013 TT power. The reasoning for the choice of cut is illustrated in Figure~\ref{fig:ax_var_z2}. We show the variance of axion fluctuations at $z=2$ for a variety of masses and $\Omega_a h^2=0.001$ (fraction $\sim 1\%$). 

For $m_a=m_{\rm lin}$ perturbations just go non-linear at $z=2$ with non-linear scale $k_{\rm nl}\approx 1 \, h\text{ Mpc}^{-1}$, while lighter ULAs are still linear at $z=2$. This suggests that non-linear effects in lensing for lighter ULAs can be safely neglected (based on the discussion of the lensing kernel in Section~\ref{observables} and in Ref.~\cite{Lewis:2006fu}). ULAs with $m_a<m_{\rm lin}$ are known, from TT anisotropies at $\ell\lesssim 10^3$ where non-linear effects are unimportant, to comprise only a sub-dominant component of the DM~\cite{Hlozek:2014lca}. We have shown that the density perturbations in such an axion should be largely unaffected by non-linearities on scales and redshift ranges relevant to CMB lensing.

Heavier ULAs with $m_a>m_{\rm lin}$ can constitute large components of the DM, and have large collapsed fractions, and thus cannot simply be ignored in the non-linear ratio. In the absence of $N$-body simulations, in order to assess the accuracy of using \textsc{halofit} for such ULAs, we compare the results of \textsc{halofit} to those of the halo model. The ULA halo model power for $m_a\geq 10^{-24}\text{ eV}$ and $\Omega_a/\Omega_d=1$ can be computed using \textsc{WarmAndFuzzy}~\cite{Marsh:2016vgj}. In order to make the comparison still fairer, we modify the halo model power spectrum, setting it strictly to linear if $\sigma^2<1$ for all $R$. The effect of non-linear modeling on the matter power and CMB lensing deflection was shown already in Figs.~\ref{fig:lensing_halofit_halomodel} and~\ref{fig:tk_halofit_halomodel}.

We illustrate the danger of using an incorrect nonlinear treatment in Figure~\ref{fig:axions_nonlinear}, where we compare an `incorrect nonlinear treatment' (na\"{i}ve use of \textsc{halofit}) to our `best approximation' for forecasts (use of $m_{\rm lin}$ and linear theory for derivatives at high $m_a$). As expected, for $m_a<m_{\rm lin}$ the non-linear modeling has no effect on the constraints. For heavier ULAs, however, the size of the error can be affected by a factor of two by incorrect non-linear modeling, and a false degeneracy direction introduced between the ULAs and the CDM content. Careful treatment of non-linear modeling is required to test the `fuzzy DM' regime with CMB-S4 lensing.

\bibliography{cmbs4_grin}
\end{document}